 \documentclass[preprint2]{aastex}

\slugcomment{Not to appear in Nonlearned J., 45.}

\shorttitle{XO3b IRAC ch2 eclipses}
\shortauthors{Morello et al.}

\begin{document}


\title{Repeatability of \textit{Spitzer}/IRAC exoplanetary eclipses with Independent Component Analysis}


\author{G. Morello, I. P. Waldmann, G. Tinetti}
\affil{Department of Physics \& Astronomy, University College London, Gower Street, WC1E6BT, UK}
\email{giuseppe.morello.11@ucl.ac.uk}



\begin{abstract}
The research of effective and reliable detrending methods for \textit{Spitzer} data is of paramount importance for the characterization of exoplanetary atmospheres. To date, the totality of exoplanetary observations in the mid- and far-infrared, at wavelengths $>$3 $\mu$m, have been taken with \textit{Spitzer}. In some cases, in the past years, repeated observations and multiple reanalyses of the same datasets led to discrepant results, raising questions about the accuracy and reproducibility of such measurements. \cite{mor14, mor15} proposed a blind-source separation method based on the Independent Component Analysis of pixel time series (pixel-ICA) to analyze IRAC data, obtaining coherent results when applied to repeated transit observations previously debated in the literature. Here we introduce a variant to pixel-ICA through the use of wavelet transform, wavelet pixel-ICA, which extends its applicability to low-S/N cases. We describe the method and discuss the results obtained over twelve eclipses of the exoplanet XO3b observed during the ``Warm \textit{Spitzer}'' era in the 4.5 $\mu$m band. The final results are reported, in part, also in \citep{ing16}, together with results obtained with other detrending methods, and over ten synthetic eclipses that were analyzed for the  ``IRAC Data Challenge 2015''.  Our results are consistent within 1 $\sigma$ with the ones reported in \cite{wong14} and with most of the results reported in \cite{ing16}, which appeared on the arXiv while this paper was under review. Based on many statistical tests discussed in \cite{ing16}, the wavelet  pixel-ICA method performs as well as or better than other state-of-art methods recently developed by other teams to analyze \textit{Spitzer}/IRAC data, and, in particular, it appears to be the most repeatable and the most reliable, while reaching the photon noise limit, at least for the particular dataset analyzed. Another strength of the ICA approach is its highest objectivity, as it does not use prior information about the instrument systematics, making it a promising method to analyze data from other observatories. The self-consistency of individual measurements of eclipse depth and phase curve slope over a span of more than three years proves the stability of Warm \textit{Spitzer}/IRAC photometry within the error bars, at the level of 1 part in 10$^4$ in stellar flux. 
\end{abstract}

\keywords{methods: data analysis - techniques: photometric - planets and satellites: atmospheres - planets and satellites: individual(XO3b)}

\section{Introduction}
Observations of transits and eclipses of exoplanets have been used, in the last decade, to characterize the nature of more than 100 of these alien worlds with unprecedented detail. Molecular, atomic and ionic signatures have been detected in the atmospheres of exoplanets through transmission spectroscopy, i.e. multiwavelength measurements of the transit depth, showing differential absorption/scatter of the stellar light through the planetary limb (e.g. \citealp{cha02, tin10, dem13}). For the brightest targets, emission spectra have been measured during (planetary) eclipses to constrain the atmospheric chemistry, pressure-temperature profile, albedo, and global circulation patterns (e.g. \citealp{cha05, swa09}). Many datasets were obtained using the InfraRed Array Camera (IRAC, \citealp{faz04}) onboard \textit{Spitzer Space Telescope}. Since depletion of the telescope's cryogen in 2009, IRAC continues to operate in the 3.6 and 4.5 $\mu$m bands, as the ``\textit{Spitzer} Warm Mission''.

A precision of 0.01$\%$ is required to study the atmospheres of giant exoplanets through transmission and/or emission spectroscopy \citep{bro01, tes13, wal15, wal15b}. Detrending instrumental systematics in raw data is necessary to achieve the target precision. It is not always obvious how to decorrelate the data using auxiliary information of the instrument and, in some cases, different methods lead to significantly different results (see e.g. \citealp{wal12, mor15}). The majority of exoplanets' transit and eclipse multi-band photometric data adopted in the literature are obtained by combining data at different wavelengths from separate epochs years apart. Repeated observations are necessary to estimate the overall level of  variability, due to astrophysical variations and possible uncorrected instrument effects. If consistent, compared to single observations, repeated observations can provide more accurate parameter values, leading to higher signal-to-noise atmospheric spectra for exoplanets, and potentially allowing the characterization of smaller exoplanets around fainter stars.

The main instrumental effect affecting \textit{Spitzer}/IRAC data at 3.6 and 4.5 $\mu$m is due to intra-pixel gain variations and spacecraft-induced motion, so-called pixel-phase effect. The measured flux from the star is correlated with its position on the detector, hence the idea of correcting the data with a polynomial function of the stellar centroid as proposed in the literature \citep{cha05, mc06, ste10, bea11}.  Multiple reanalyses of the same datasets with the polynomial method show that, in some cases, results can be sensitive to some specific options/variants, such as the degree of the polynomial adopted, partial data rejection, including temporal or other decorrelations (e.g. \citealp{bea08, des09, ste10, bea11, knu11}). Recently, several alternative methods have been proposed to decorrelate \textit{Spitzer}/IRAC data: gain mapping \citep{bal10, cow12, knu12, lew13, zel14}, bilinearly interpolated sub-pixel sensitivity mapping (BLISS, \citealp{ste12}), Independent Component Analysis using pixel time series (pixel-ICA, \citealp{mor14, mor15}), pixel-level decorrelation (PLD, \citealp{dem15}), and Gaussian Process models \citep{eva15}. A comparison between pixel-ICA and PLD methods is reported in \cite{mor15b}. The discussion about the detrending methods, their performances, reliability and potential biases for \textit{Spitzer}/IRAC data is a hot topic. In the ``IRAC Data Challenge 2015'' different methods have been tested over synthetic data created by the IRAC team, which contain ten simulated eclipse observations of the exoplanet XO3b \citep{ing16}. Researchers were also encouraged to reanalyze a similar set of real observations obtained in the 4.5 $\mu$m band.

In this paper, we describe an evolution of the pixel-ICA method proposed in \cite{mor14, mor15, mor15b}, and present the results obtained by applying said method to the analysis of twelve eclipses of the exoplanet XO3b taken with Warm \textit{Spitzer}/IRAC in the 4.5 $\mu$m band.
Pixel-ICA method differs from the other detrending methods proposed in the literature, as it is an unsupervised machine learning algorithm. The lack of any prior assumptions about the instrument systematics and astrophysical signals ensures a high degree of objectivity, and indicates that the same method could be applied to detrend data taken with different instruments. Pixel-ICA method gave coherent results when applied over multiple transit observations of the exoplanets HD189733b and GJ436b, for which the previous literature reported discrepant results \citep{mor14, mor15}, and over simulated observations with a variety of instrumental systematics \citep{mor15b}. Similar techniques have been used in the literature to detrend \textit{Spitzer}/IRS and \textit{Hubble}/NICMOS data, the main difference being in the choice of the input time series \citep{wal12, wal14, wal13}.
The ability of ICA to decorrelate non-gaussian signals is inherently limited to a low gaussian white noise amplitude relative to the non-gaussian signals. In this paper, we propose a wavelet pixel-ICA algorithm, which outperforms the traditional pixel-ICA algorithm in low-S/N observations, extending applicability to planetary eclipses taken during the ``Warm \textit{Spitzer}'' era.

XO3b is a hot Jupiter ($M_p = $11.7 $\pm$ 0.5 $M_{Jup}$, \citealp{jk08, hir11}) in an eccentric orbit ($e = $0.283 $\pm$ 0.003, \citealp{knu14}) with a period of 3.19 days and orbital semimajor axis of $a = 0.045$ AU \citep{winn08}. The host star is F5V with $T_{*} =$ 6760 $\pm$ 80 K, and $\log{g} = $4.24 $\pm$ 0.03 \citep{winn08, tor12}. A previous analysis of the 12 eclipses reported an average eclipse depth of 1.58$_{-0.04}^{+0.03} \times$ 10$^{-3}$ relative to the out-of-eclipse flux (star+planet), and a phase curve slope of 6.0$_{-1.6}^{+1.3} \times$ 10$^{-4}$ days$^{-1}$ \citep{wong14}.  Here we compare our results with the ones reported in \cite{wong14}. 

\section{Data Analysis}
\subsection{Observations}
We analyze twelve eclipse observations of XO3b taken with \textit{Spitzer}/IRAC in the 4.5 $\mu$m band. Ten individual eclipses were observed over 6 months (Nov 11, 2012 - May 24,  2013), including two sets of three consecutive eclipses, another eclipse is contained within a full-orbit observation on May 5, 2013 (PID:  90032). Each individual observation consists of 14,912 frames over 8.4 hr using IRAC's sub-array readout mode with 2.0 s integration time. In sub-array mode 64 frames are taken consecutively, the reset time is $\sim$1 s. We extracted 14,912 frames from the full-orbit observation to analyze the light-curve of the eclipse over a time interval similar to other observations. The last eclipse was extracted out of a 66 hr observation on April 8, 2010 (PID: 60058). Table \ref{tab1} reports the dates in which the eclipses were observed.
\begin{table}[!h]
\begin{center}
\caption{Eclipse observations dates and orbit numbers of XO3b. \label{tab1}}
\begin{tabular}{ccccc}
\tableline\tableline
Obs. Number & UT Date & Orbit Number\\
\tableline
1 & 2010 Apr 8 & 0\\
2 & 2012 Nov 11 & 297\\
3 & 2012 Nov 17 & 299\\
4 &  2012 Nov 20 & 300\\
5 &  2012 Nov 23 & 301\\
6 & 2012 Dec 2 & 304\\
7 & 2012 Dec 9 & 306\\
8 & 2013 Apr 22 & 348\\
9 & 2013 May 5 & 352\\
10 & 2013 May 18 & 356\\
11 & 2013 May 21 & 357\\
12 & 2013 May 24 & 358\\
\tableline
\end{tabular}
\end{center}
\end{table}

\subsection{The eclipse model}
\begin{figure}[!h]
\plotone{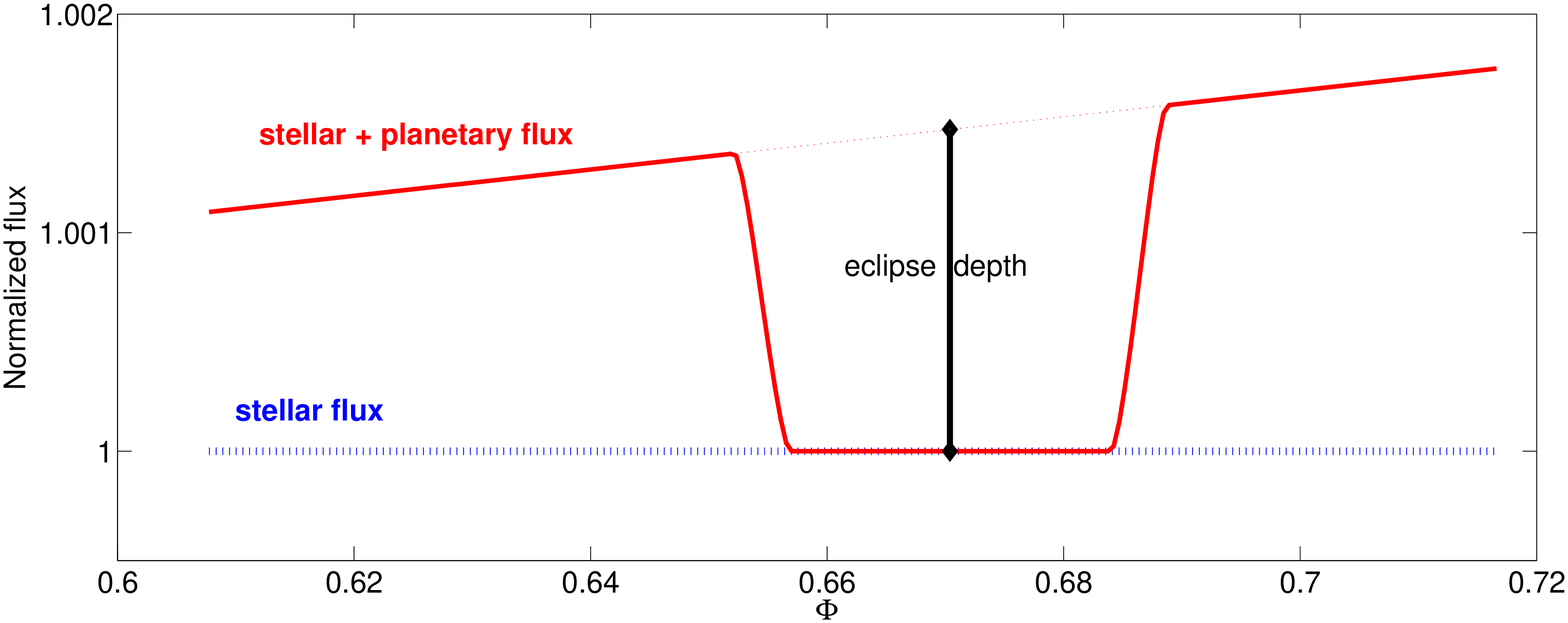}
\caption{Scheme of the eclipse model adopted: the stellar flux is constant and normalized to 1 (blue dashed line), the planetary flux is a linear function of time, and disappears during the eclipse (red line). The eclipse depth is the extrapolated planetary flux, in units of stellar flux, at the eclipse center. \label{fig1}}
\end{figure}

In our model the stellar flux is constant in time, and normalized to 1. We adopt the formalism of \cite{ma02} for the eclipse model, accounting for the planet being occulted by the star. We approximate the planet's phase curve in the region of the eclipse as a linear function of the time, the slope is called ``phase constant'', adopting the same terminology of \cite{wong14}. The slope is only due to planet's flux variations. While the planet is completely occulted by the star, the flux is constantly 1. The eclipse depth is defined as the (unseen) planet's flux at the center of eclipse, in units of stellar flux (see Figure \ref{fig1}). When fitting the eclipse models, the orbital parameters are fixed to the values reported in Table \ref{tab2}, taken from \cite{wong14}, while the center of eclipse, eclipse depth and phase constant are free parameters to determine. We also considered models with zero phase curve's slope (see Appendix \ref{app_zeroslope}).
\begin{table}[!h]
\begin{center}
\caption{Values of the parameters fixed while generating the eclipse models. \label{tab2}}
\begin{tabular}{cc}
\tableline\tableline
orbital period, $P$ (days) & 3.19153285\\
scaled semi-major axis, $a/R_*$ & 7.052\\
inclination, $i$ (deg) & 84.11\\
eccentrity, $e$ & 0.2833\\
argument of periastron, $\omega$ (deg) & 346.8\\
\tableline
\end{tabular}
\end{center}
\end{table}

\subsection{Wavelet ICA}
\subsubsection{Continuous Wavelet Transform (CWT)}
The wavelet transform (WT) decomposes a given signal, $x(t)$, into its frequency components. This is done by convolving the time signal with a basis of highly localized impulses or ``wavelets''. To fix the ideas, we assume that $x(t)$ is a time series, although this is not necessary, as the DWT can be applied to different kinds of signals. The individual wavelet functions are derived from a single ``mother wavelet'', $\psi(t)$, through translation and dilation of the mother wavelet. The mathematical definition of the CWT is 
\begin{equation}
c_{\tau,\varphi} = \int_{\mathbb{R}} x(t) \psi_{\tau,\varphi}(t) \ dt
\end{equation}
\begin{equation}
\psi_{\tau,\varphi}(t) = \frac{1}{\sqrt{2}} \ \psi \left ( \frac{t-\tau}{\varphi} \right )
\end{equation}
where $\psi_{\tau,\varphi}$ is the mother wavelet for a given scaling $\varphi$ and translation $\tau$, and $c_{\tau,\varphi}$ is the wavelet coefficient with respect to $\tau$ and $\varphi$.\\
If the wavelet basis is orthogonal, the inverse wavelet transform can be used to reconstruct the original time series:
\begin{equation}
x(t) = \sum_{\varphi \in \mathbb{Z}} \sum_{\tau \in \mathbb{Z}} c_{\tau,\varphi} \psi_{\tau,\varphi}(t)
\end{equation}
The mother wavelet can be chosen among a variety of wavelet families with different properties. For more details we refer to the relevant literature, e.g. \cite{dau92, pw00}.

\subsubsection{Discrete Wavelet Transform (DWT)}
\label{DWT}
Astronomical data are usually in the form of discrete time series. For the DWT the mother wavelet is denoted by $h(t)$, and the scaling function, also called ``father wavelet'', is denoted by $g(t)$. The mother and father wavelets act as high-pass and low-pass frequency filters, respectively. They are related by
\begin{equation}
g(L-1-t) = (-1)^t \ h(t),
\end{equation}
where $L$ is the filter length and corresponds to the number of points in the time series $x(t)$.\\
The one-level DWT is defined by
\begin{equation}
\label{eqn_A1}
cA_{1}(\tau) = (x * g)(t) \downarrow 2 = \sum_{t = - \infty}^{+ \infty} x(t) \ g(2 \tau -t)
\end{equation}
\begin{equation}
\label{eqn_D1}
cD_{1}(\tau) = (x * h)(t) \downarrow 2 = \sum_{t = - \infty}^{+ \infty} x(t) \ h(2 \tau -t)
\end{equation}
The $cA_{1}$ time series approximates the underlying low-frequency trend of $x(t)$ (average coefficients), while the $cD_{1}$ time series represents a higher frequency component (detail coefficients). They are down-sampled by a factor of 2 (``$\downarrow 2$'' in Equations \ref{eqn_A1} and \ref{eqn_D1}) with respect to the original time series, because of the Nyquist theorem.\\
It is possible to apply the $g$ and $h$ filters to the $cA_{1}$ time series, then obtaining new sets of coefficients, $cA_{2}$ and $cD_{2}$, and iterate the process. The n-level DWT includes the $cA_{n}$ series of average coefficients, down-sampled by a factor of 2$^n$, and n series $cD_{1}$-$cD_{n}$ of detail coefficients, representing bands of higher frequencies.
The original data can be reconstructed by reversing the process:
\begin{equation}
\label{invDWT}
x(t) = cA_n(\tau) \ g(-t + 2 \tau ) + \sum_{i=1}^{n} \sum_{\tau = - \infty}^{+ \infty} (cD_i \ h(-t + 2 \tau ))
\end{equation}

\subsubsection{Wavelet ICA}
In this section we describe the wavelet ICA algorithm, which is used in a variety of contexts, such as medical sciences (e.g. \citealp{laf06, inu07, mam12}), engineering (e.g. \citealp{lz05}), acoustic (e.g. \citealp{mou06, zhao06}), image denoising (e.g. \citealp{kt14}), and astrophysics (e.g. \citealp{wal14}).

Be $\textbf{x} = (x_1, x_2, \dots, x_m)^T$ the column vector of observed signals. In this paper, $\textbf{x}_k$ are individual pixel time series, so-called pixel-light-curves, which are mixtures of different source signals, astrophysical or instrumental in nature, and gaussian noise. The formalisms adopted in this subsection is valid in a more general context, where the $x_k$ can be any kind of mixed signals. ICA is a linear transformation of the observed (mixed) signals which minimizes the mutual information to decorrelate the independent components:
\begin{equation}
\textbf{x} = \textbf{A} \textbf{s}, \ \textbf{s} = \textbf{W} \textbf{x}
\end{equation}
where $\textbf{s} = (s_1, s_2, \dots, s_m)^T$ is the column vector of the original source signals, $\textbf{A}$ is the matrix of mixing coefficients, and $\textbf{W} = \textbf{A}^{-1}$. We refer the reader to \cite{wal12, mor14, mor15} for additional details.

The ability of ICA to decorrelate non-gaussian signals is inherently limited to a low gaussian noise amplitude relative to the non-gaussian signals. Wavelet ICA algorithms are designed to be less sensitive to white noise compared to the simple ICA separation, described above.
In wavelet ICA algorithms, the DWT is applied to the observed signals:
\begin{equation}
x_k(t) \rightarrow \hat{x}_k = (cA_{k,n}, cD_{k,n}, \dots, cD_{k,1})
\end{equation}
\begin{equation}
\textbf{x}(t) \rightarrow \hat{\textbf{x}} = (\hat{x}_1(t), \hat{x}_2(t), \dots, \hat{x}_m(t))^T
\end{equation}
The ICA separation is performed with the series of coefficients:
\begin{equation}
\label{eqn:waveletICA}
\hat{\textbf{s}} = \hat{\textbf{W}} \hat{\textbf{x}}
\end{equation}
The independent components series of coefficients are:
\begin{equation}
\hat{s}_l = (cA_{l,n}, cD_{l,n}, \dots, cD_{l,1})
\end{equation}
They can be converted into the time domain through inverse DWT (Equation \ref{invDWT}).

The DWT preliminarly separates the high-frequency components from the low-frequency trend, enhancing the ability of ICA to disentangle the low-frequency independent components. This step is particularly important in cases where the gaussian noise is dominant. Additional processing options/variants have been proposed in the literature to further improve the ICA performances in specific contexts, e.g. coefficients' thresholding \citep{ste81,don95}, suppression of some frequency ranges \citep{lz05}, taking individual levels as input to ICA \citep{inu07,mam12}. In this paper, we aim to provide the most objective analysis of the datasets, with minimal prior assumptions, hence those variants are not considered. The impact of those variants and other operations to the data will be carefully investigated in future studies.

\subsection{Detrending method, light-curve fitting and error bars}

In this Section we list the main steps of the wavelet pixel-ICA method, followed by a more accurate description and comments:
\begin{enumerate}
\item Selecting an array of pixels. The raw light-curve is the sum of the individual pixel time series within the selected array.
\item Removing outliers.
\item Subtracting the background from the raw light-curve.
\item Computing the wavelet transforms of the time series from the pixels within the selected array, hereafter called pixel-light-curves.
\item Performing ICA decomposition of the wavelet-transformed pixel-light-curves.
\item Computing the inverse wavelet transforms of the independent components.
\item Simultaneous fitting of the components (except the eclipse one) and astrophysical model on the raw light-curve.
\item Estimating parameter error bars.
\end{enumerate}

\subsubsection{Selecting the pixel-array}
We use squared arrays of pixels as photometric apertures; in this paper, we tested 5$\times$5 and 7$\times$7 arrays with the stellar centroid at their centers. By default, all pixel-light-curves within the selected array are also used to decorrelate the instrument systematics through ICA. Our previous analyses of transit observations indicated the 5$\times$5 and 7$\times$7 arrays to be optimal choices, and, in general, results were very little affected by the choice of different arrays \citep{mor14, mor15}.

\subsubsection{Outlier rejection}
We flag and correct outliers in the flux time series. First, we calculate the standard deviations of any set of five consecutive points and take the median value as the representative standard deviation. We define the expected value in one point as the median of the four closest points, i.e. two before and two after. Points differing  from their expected values by more than five times the standard deviation are flagged as outliers. They are then replaced with the mean value of the points immediately before and after, or, in case of two consecutive outliers, with a linear interpolation between the closest points which are not outliers. We checked that the outliers removed after this process are coincident with outliers that would have been spotted by eye. They are less than 0.35$\%$ the number of points in each observation.

\subsubsection{Background subtraction}
\label{back}
The background is estimated by taking the mean flux over four arrays of pixels with the same size of the selected array (5$\times$5 or 7$\times$7) near the corners of the sub-array area. In Appendix \ref{app_back} we discuss how this preliminary step improves the results. Here we anticipate that the impact on individual measurements of the eclipse depth is at the level of $\sim$10$^{-5}$, well below the error bars. However, this difference might become significant when averaging results from multiple observations, and the error bars are reduced. 

\subsubsection{Wavelet transforms}
The main novelty of the algorithm proposed in this paper is that the pixel-light-curves are wavelet-transformed before performing the ICA separation. More specifically, we adopt one-level DWT with mother wavelet Daubechies-4 \citep{dau92}. We found that different choices of the mother wavelet, among different families and numbers, lead to exactly the same results, and higher-level DWTs are not useful. We also investigated the effect of binning the time series (see Appendix \ref{app_bin}).

\subsubsection{ICA decomposition}
It is performed on the wavelet-transformed pixel-light-curves (see Equation \ref{eqn:waveletICA}).

\subsubsection{Inverse wavelet transforms}
The independent components are transformed in the time domain through inverse DWTs (see Equation \ref{invDWT}).

\subsubsection{MCMC fitting}
\label{MCMC}
The raw light-curves are linear combinations of the independent components. One of the components is the eclipse signal (with some residual systematics), other components may be instrumental systematics and/or other astrophysical signals. Instead of fitting an eclipse model to the eclipse component, more robust estimates of the eclipse parameters are obtained by fitting a linear combination of the eclipse model and the non-eclipse components, hereafter full models, to the raw light-curves. The free parameters of the eclipse model are fitted together with the scaling coefficients of the independent components. First estimates of the parameters and scaling coefficients are obtained through a Nelder-Mead optimization algorithm \citep{lag98}; they are then used as optimal starting points for an Adaptive Metropolis algorithm with delayed rejection \citep{haa06}, generating chains of 300,000 values. The output chains are samples of the posterior (gaussian) distributions. We adopt the mean values of the chains as final best estimates of the parameters, and the standard deviations as zero-order error bars, $\sigma_{par,0}$.

\subsubsection{Final error bars}
\label{error_bars}
The final parameter error bars are:
\begin{equation}
\label{eqn:sigmapar}
\sigma_{par} = \sigma_{par,0} \sqrt{ \frac{ \sigma_{0}^2 + \sigma_{ICA}^2 }{ \sigma_{0}^2}}
\label{eqn:sigma_par}
\end{equation}
$\sigma_{0}^2$ is the sampled likelihood variance, approximately equal to the variance of the residuals for the best transit model; $\sigma_{ICA}^2$ is a term accounting for the potential bias  of the components obtained with ICA.
\begin{equation}
\label{eqn:sigmaica}
\sigma_{ICA}^2 = \sum_{j} o_j^2 \textbf{ISR}_j
\end{equation}
where $\textbf{ISR}$ is the so-called Interference-to-Signal-Ratio matrix, and $o_j$ are the coefficients of the non-eclipse components. The term relative to the goodness of the fit of the components does not appear in Equation \ref{eqn:sigmaica}, as it is automatically included in $\sigma_0$ when the components' coefficients and astrophysical parameters are fitted simultaneously.

\subsection{Results}

\begin{figure*}
\epsscale{1.0}
\hspace{-1cm}
\plotone{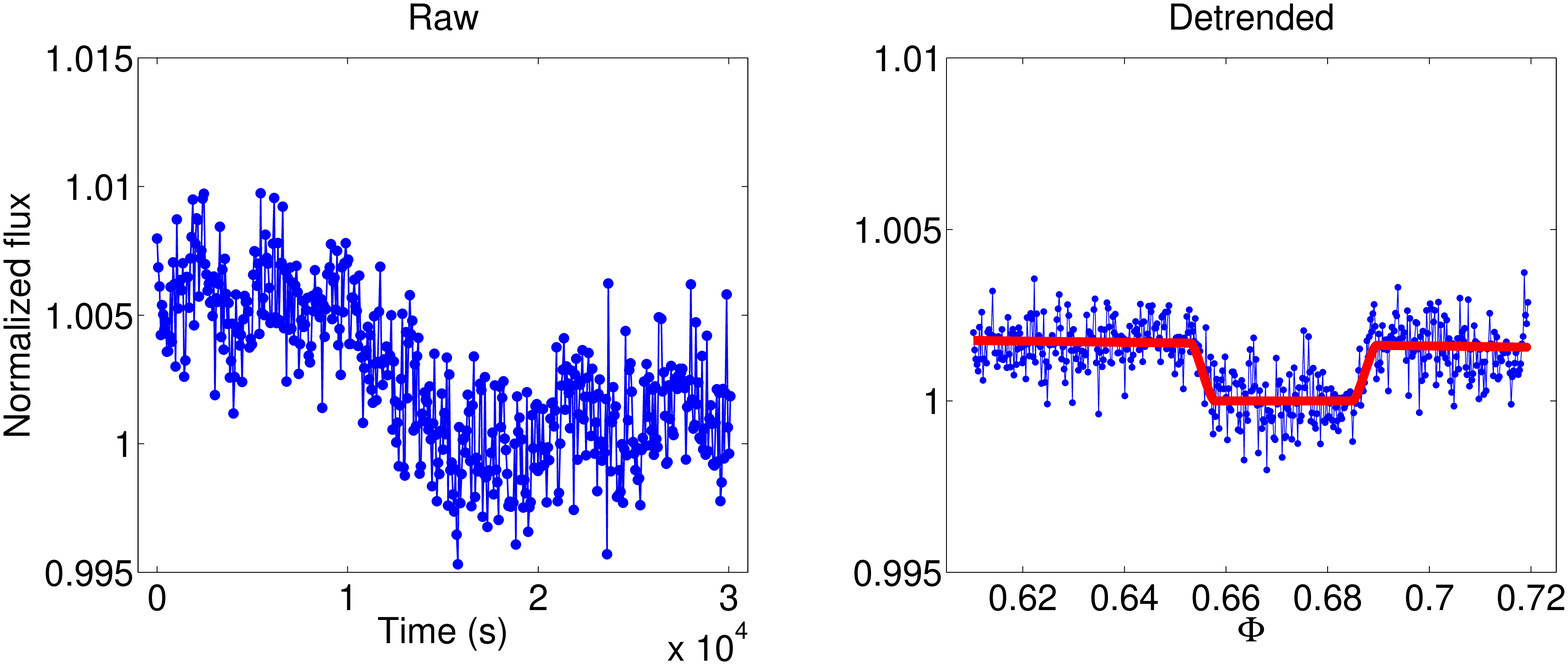}
\hspace{-1cm}
\plotone{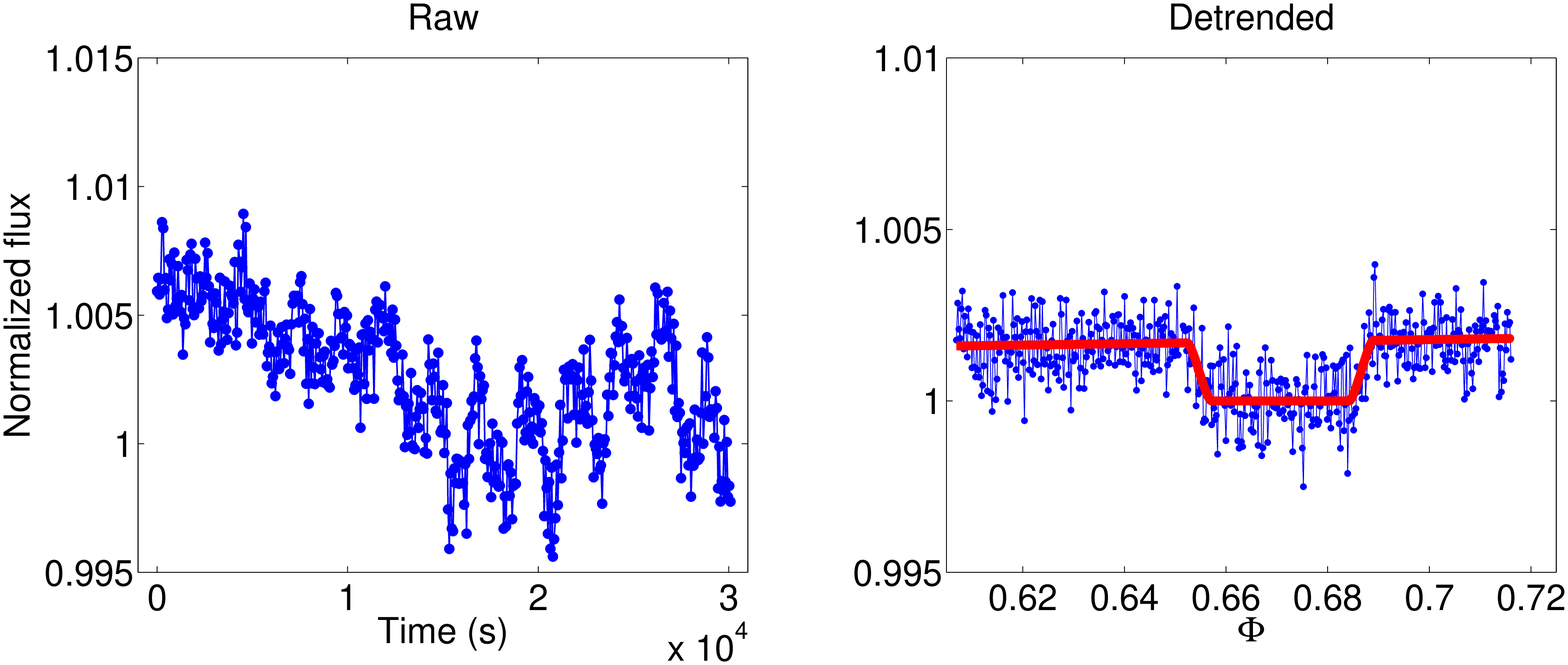}
\\
\hspace{-1cm}
\plotone{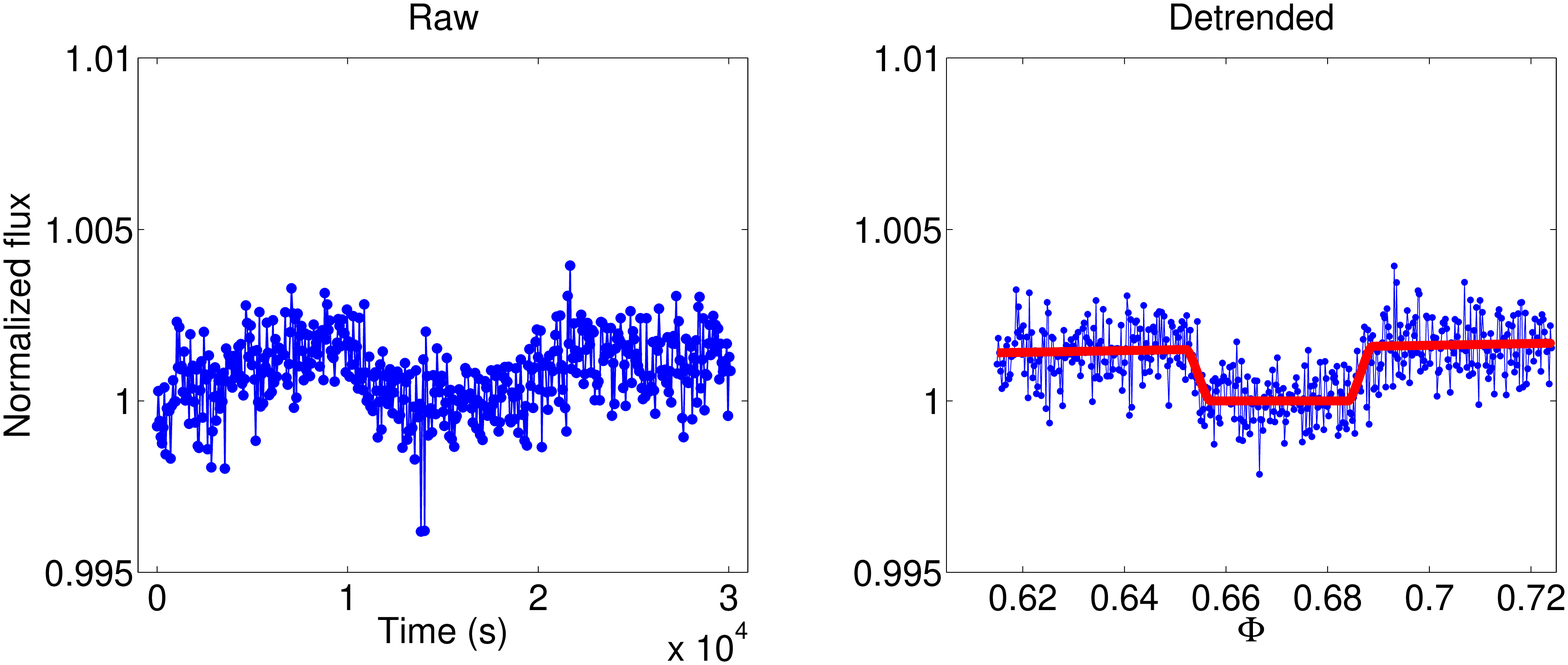}
\hspace{-1cm}
\plotone{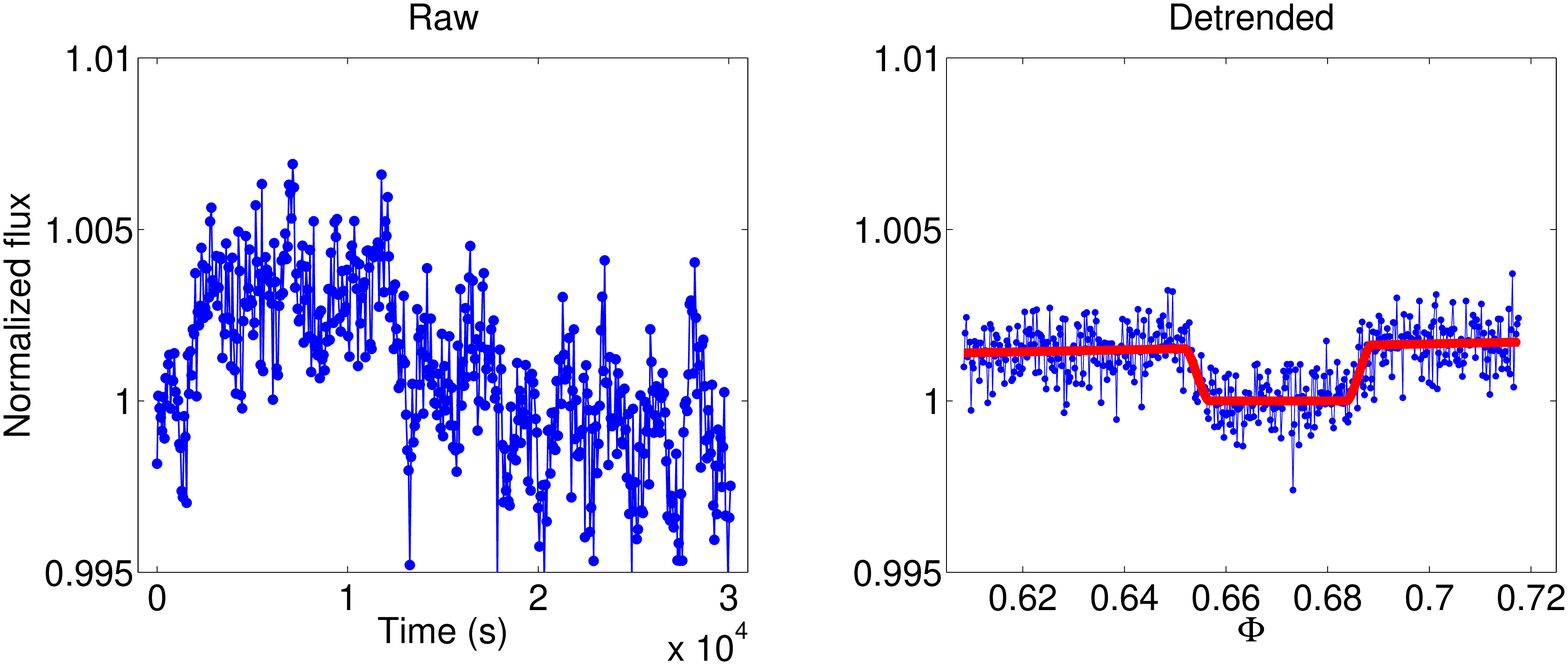}
\\
\hspace{-1cm}
\plotone{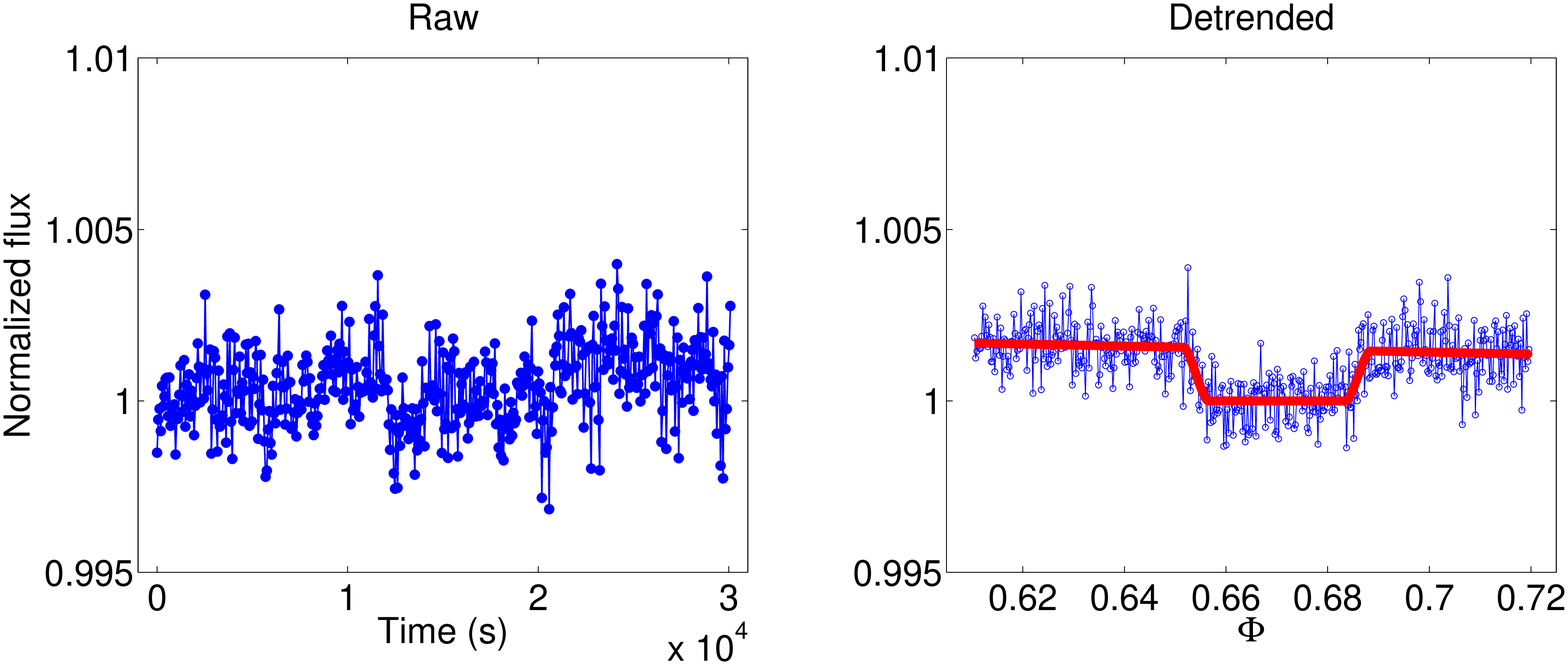}
\hspace{-1cm}
\plotone{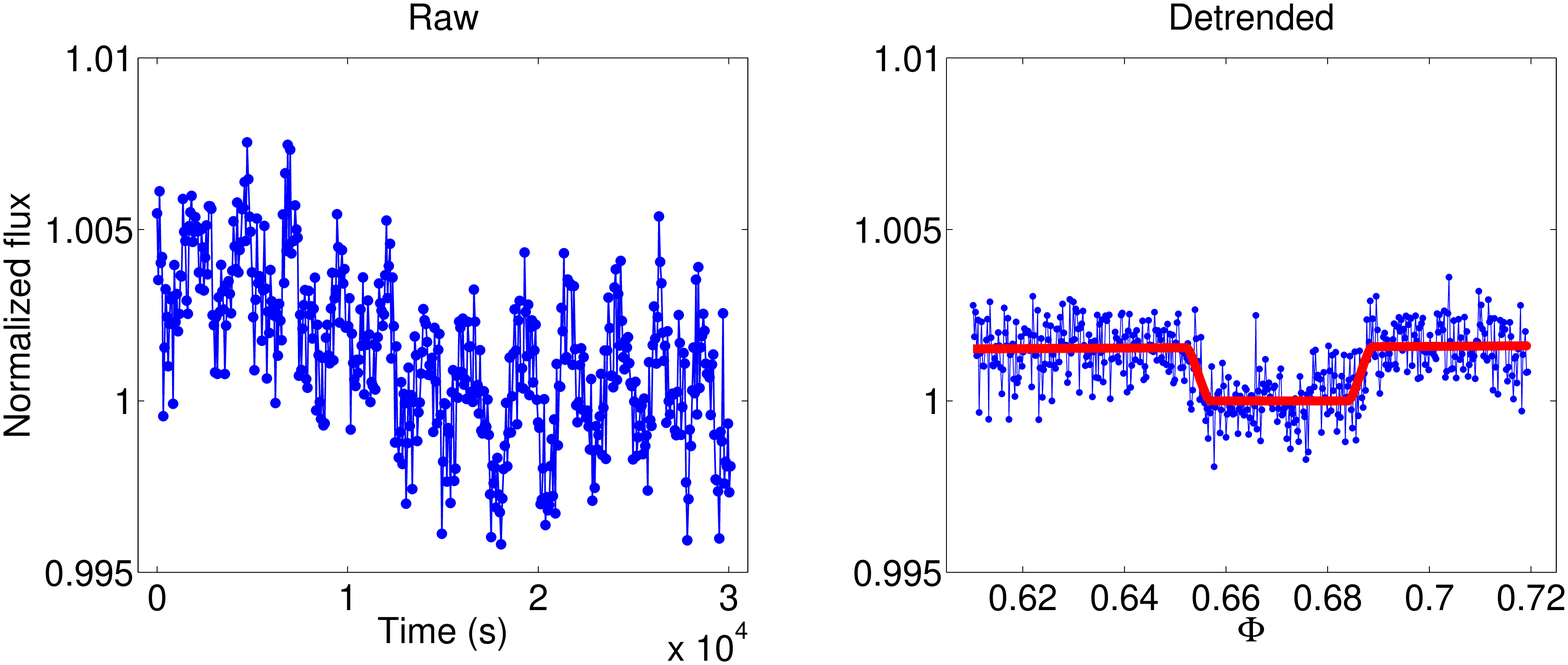}
\\
\hspace{-1cm}
\plotone{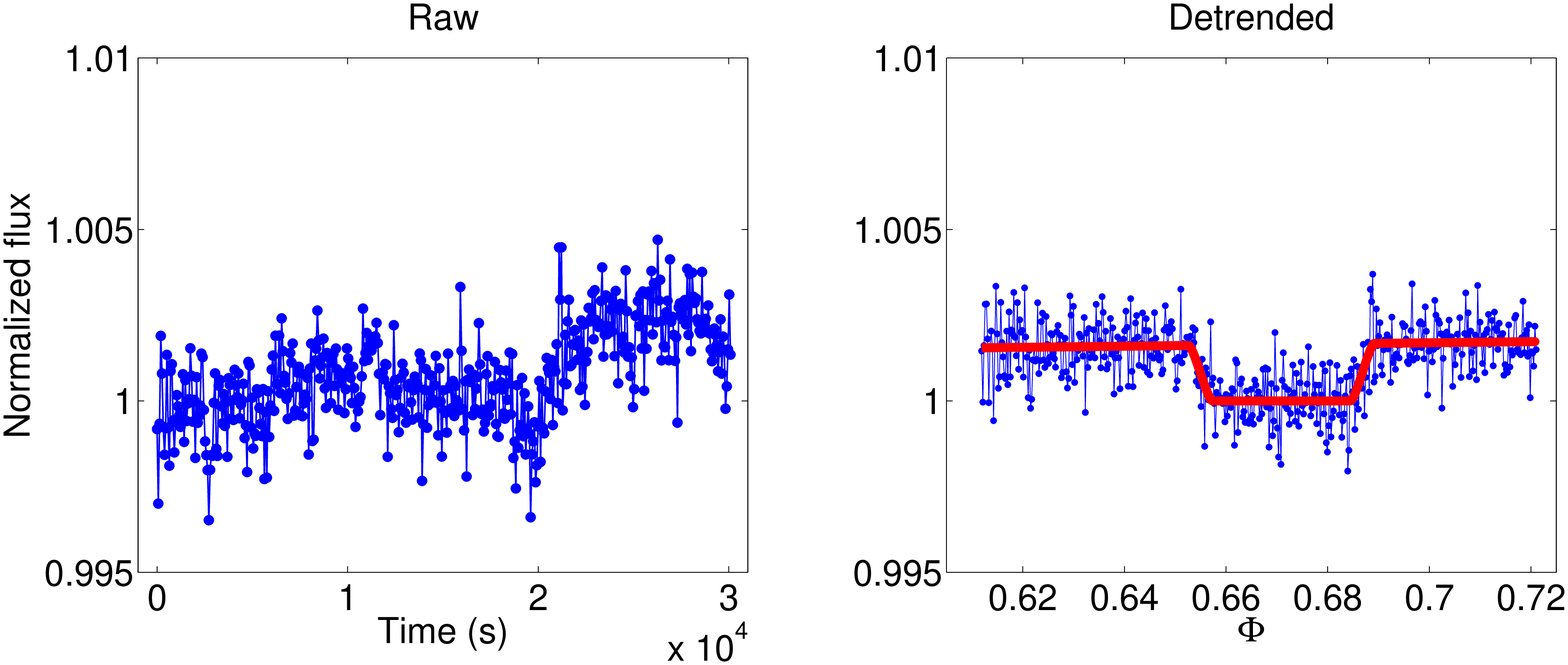}
\hspace{-1cm}
\plotone{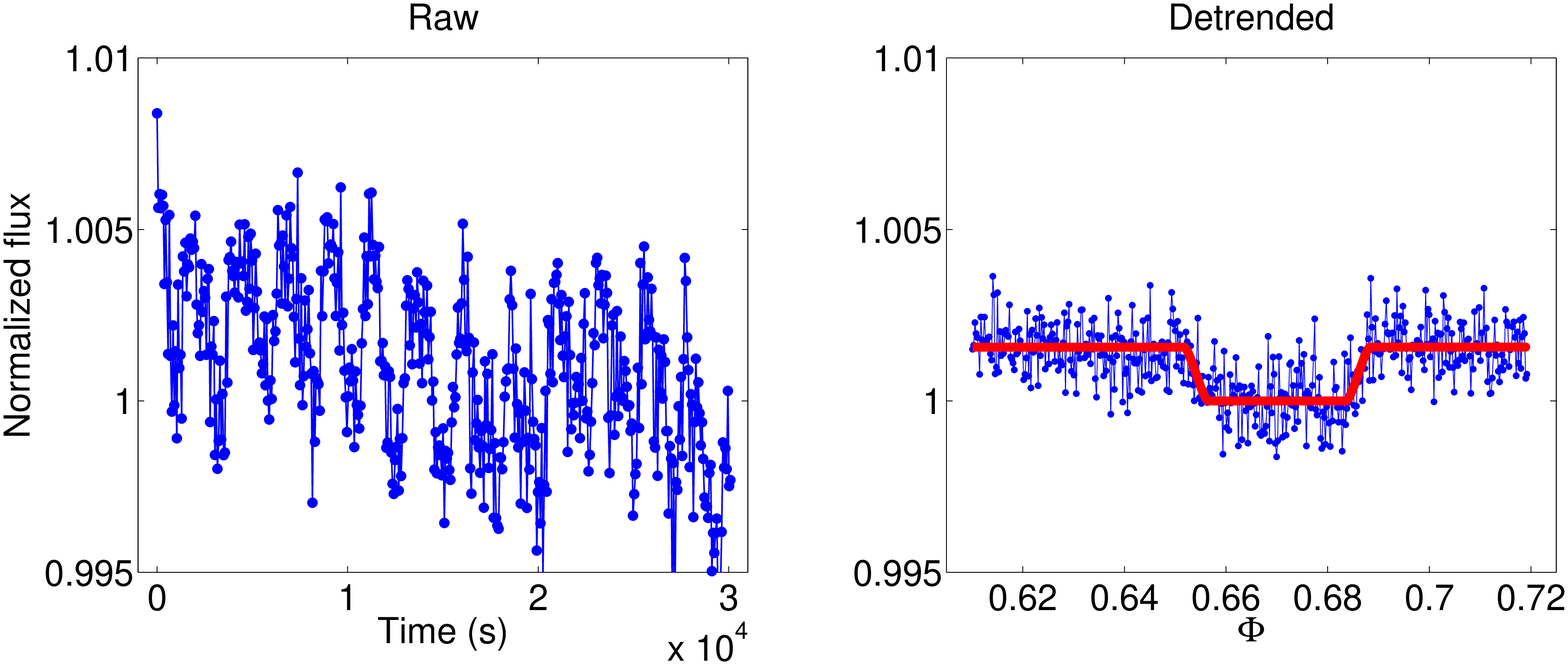}
\\
\hspace{-1cm}
\plotone{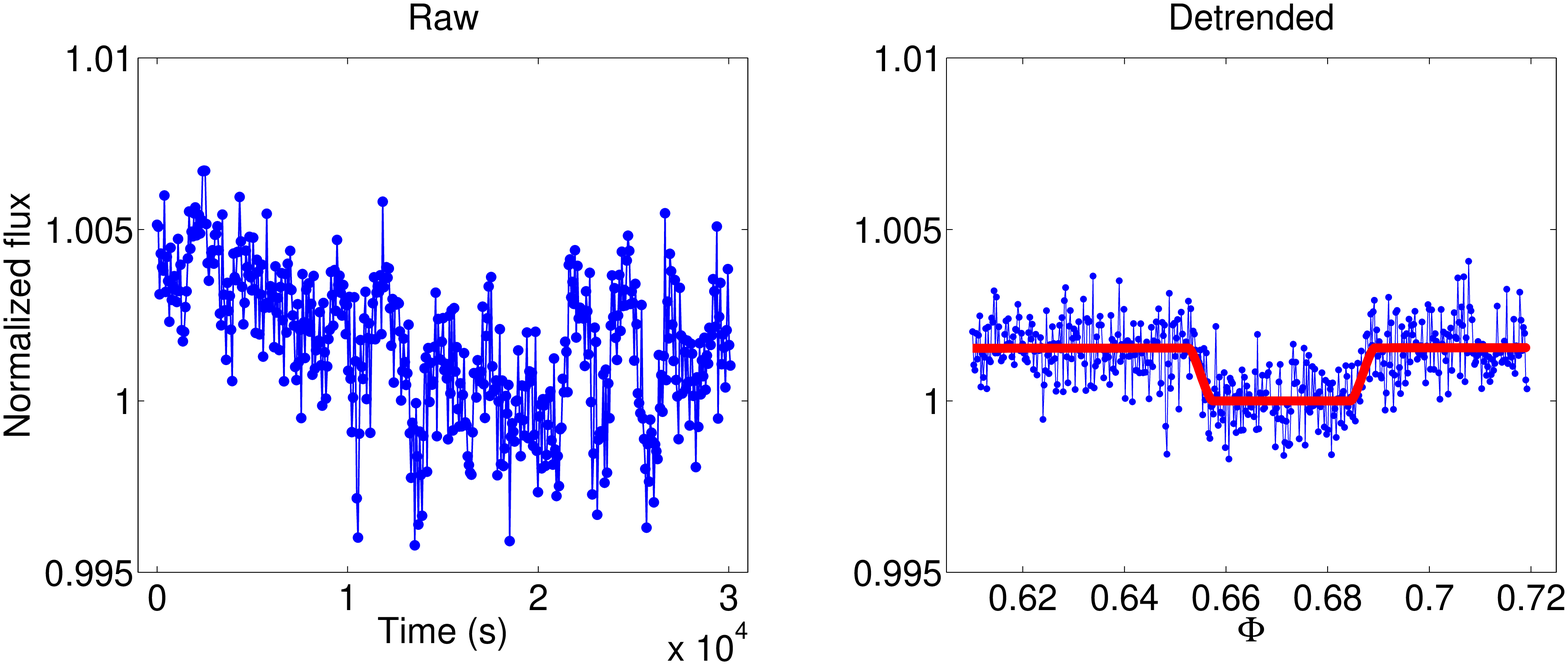}
\hspace{-1cm}
\plotone{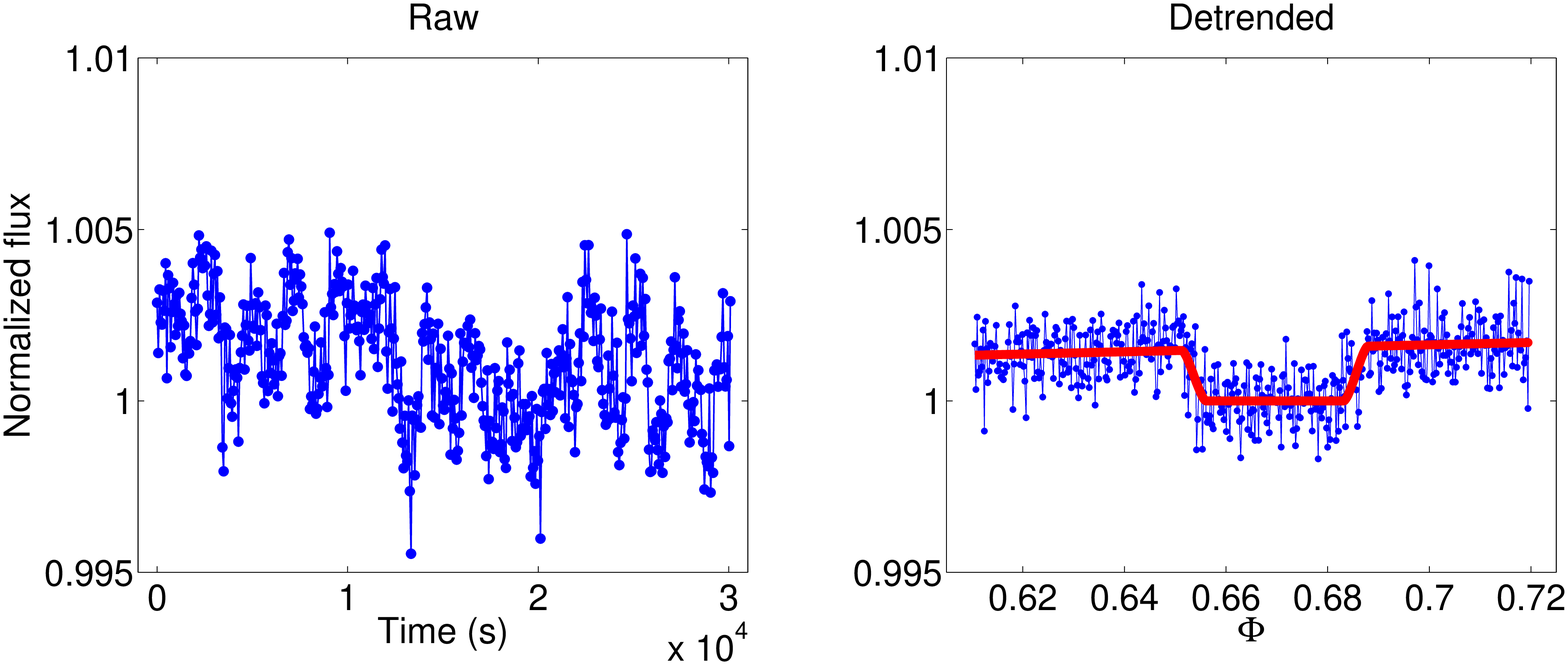}
\\
\hspace{-1cm}
\plotone{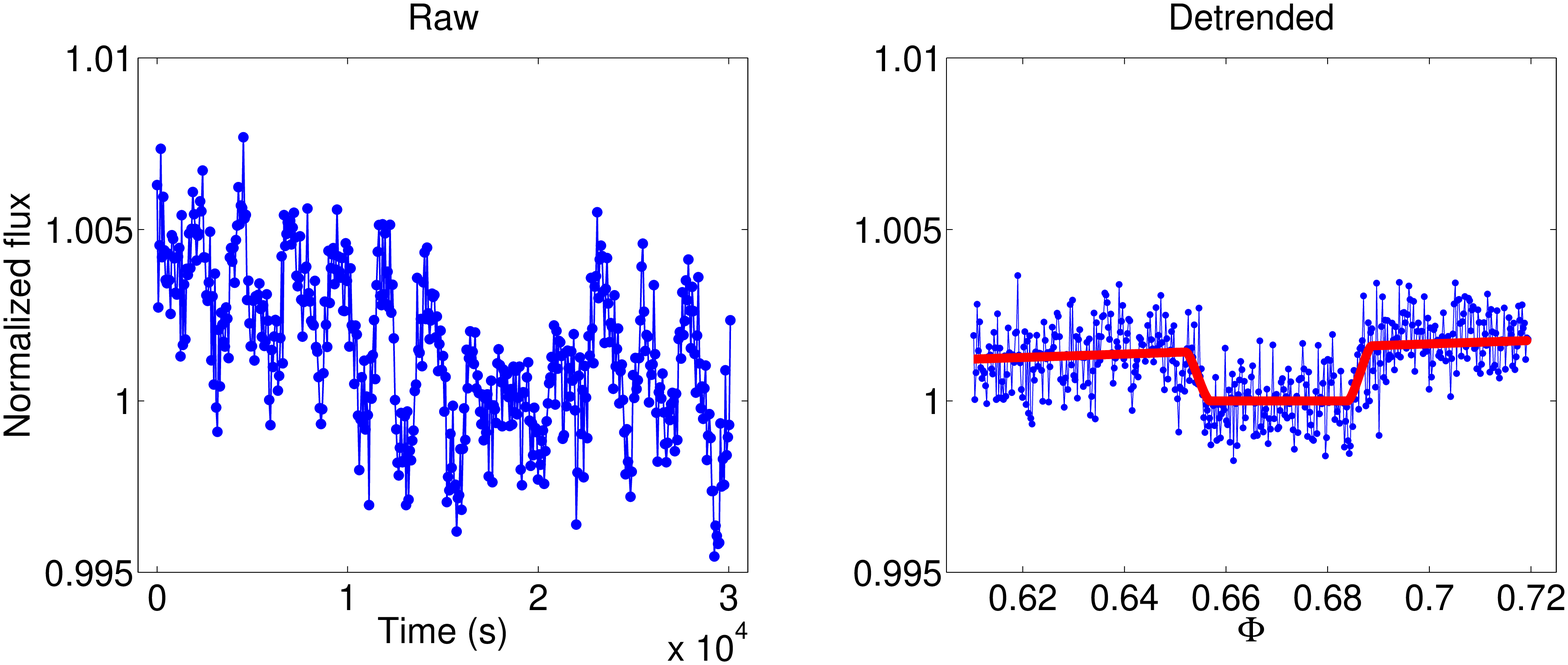}
\hspace{-1cm}
\plotone{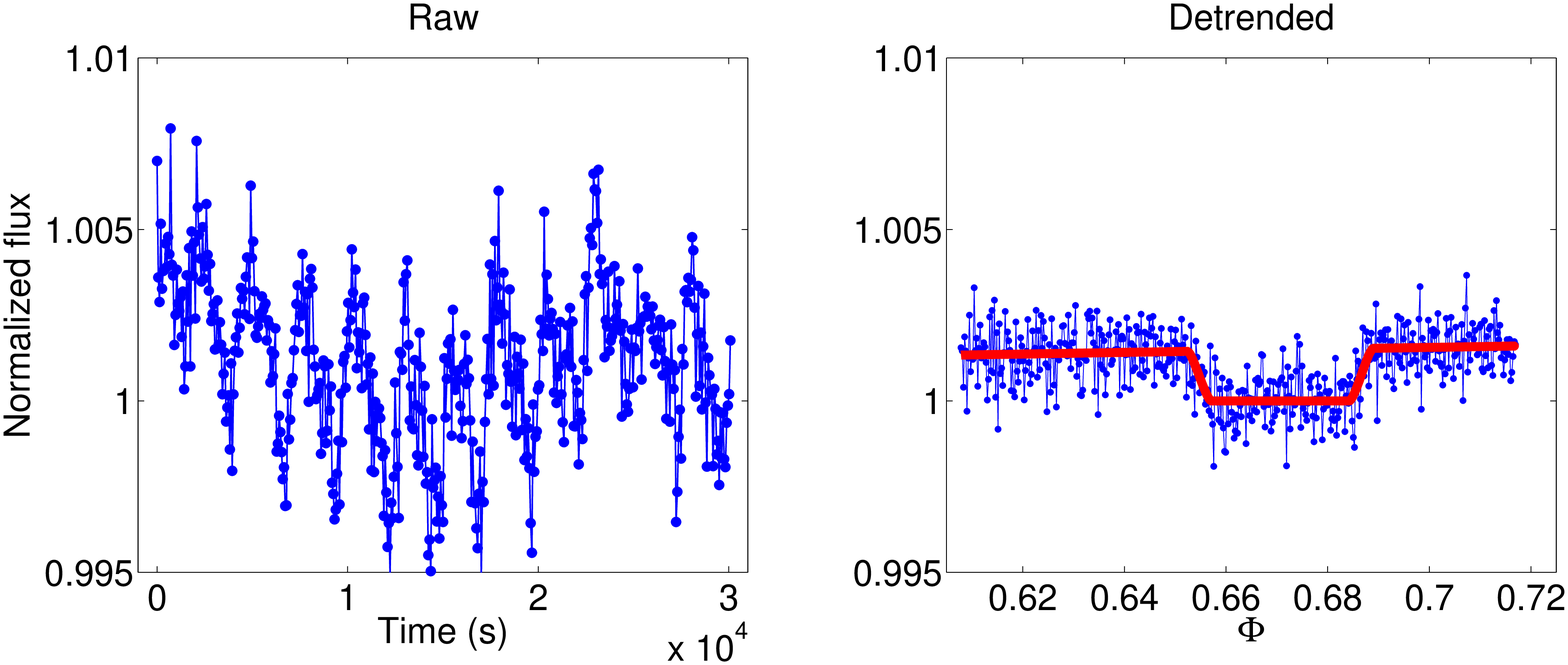}
\caption{Left panels: (blue) raw light-curves obtained from 5$\times$5 array of pixels. Right panels: (blue) detrended eclipse light-curves obtained with wavelet pixel-ICA method, and (red) best eclipse models.  All the light-curves are binned over 32 frames, i.e. $\sim$64 s. \label{fig2}}
\end{figure*}

\begin{figure*}
\epsscale{2.0}
\plotone{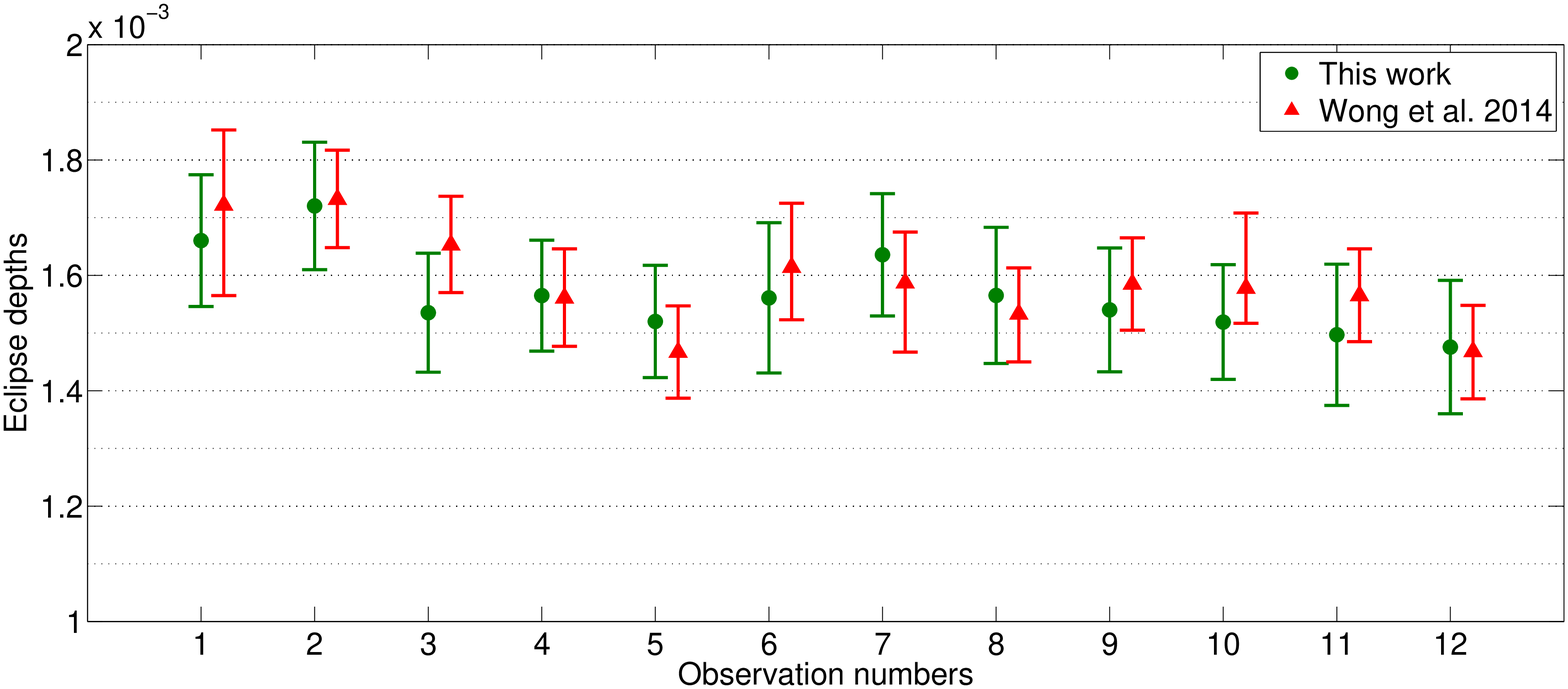}
\plotone{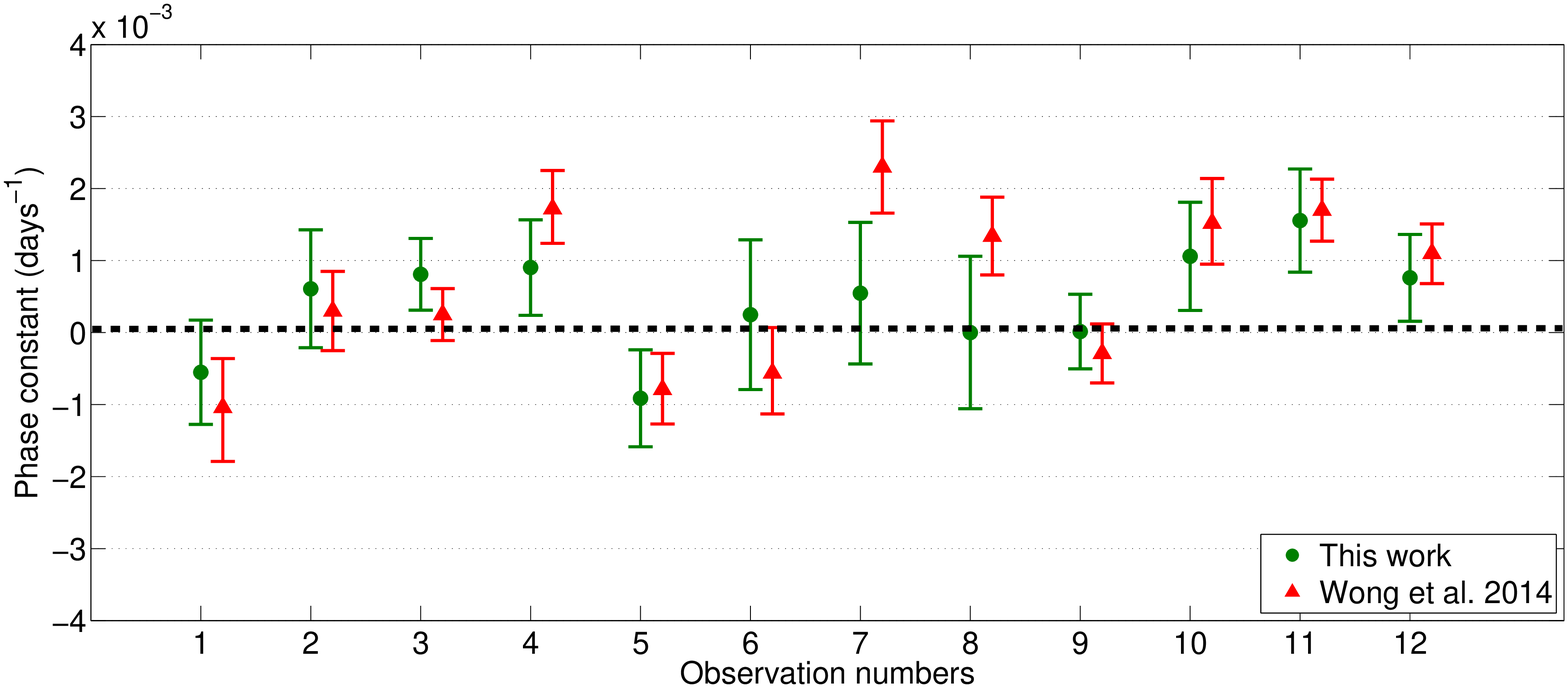}
\caption{Top panel: (green circles) individual best eclipse depth measurements obtained in this work , and (red triangles) results from \cite{wong14}. Bottom panel: the same for individual measurements of the phase constant.\label{fig3}}
\end{figure*}

Figure \ref{fig2} shows the raw light-curves for the twelve observations, the correspondent detrended eclipses and best models, obtained using the 5$\times$5 array and binning over 32 frames, i.e. $\sim$64 s. The individual measurements of eclipse depth and phase constant are reported in Figure \ref{fig3} and Table \ref{tab3}. 
\begin{table*}[!h]
\begin{center}
\caption{Individual best parameters results obtained in this work. \label{tab3}}
\begin{tabular}{ccccc}
\tableline\tableline
Obs. Number & Depth (10$^{-3}$) & 1 $\sigma$ error & Phase constant (10$^{-4}$ days$^{-1}$) & 1$\sigma$ error\\
\tableline
1 & 1.66 & 0.11 & -6 & 7\\
2 & 1.72 & 0.11 & 6 & 8\\
3 & 1.54 & 0.10 & 8 & 5\\
4 & 1.56 & 0.10 & 9 & 7\\
5 & 1.52 & 0.10 & -9 & 7\\
6 & 1.56 & 0.13 & 2 & 10\\
7 & 1.64 & 0.11 & 5 & 10\\
8 & 1.57 & 0.12 & 0 & 11\\
9 & 1.54 & 0.11 & 0 & 5\\
10 & 1.52 & 0.10 & 11 & 8\\
11 & 1.50 & 0.12 & 16 & 7\\
12 & 1.48 & 0.12 & 8 & 6\\
\tableline
\end{tabular}
\end{center}
\end{table*}
The results from all epochs are consistent within the error bars, suggesting the lack of any detectable astrophysical variability for this system, and residual instrument variability. By taking the weighted means of the individual measurements, we obtain global best estimates of (1.57 $\pm$ 0.03)$\times$10$^{-3}$ for the eclipse depth, and (4.4 $\pm$ 2.0)$\times$10$^{-4}$ days$^{-1}$ for the phase constant.
\begin{figure*}[!h]
\epsscale{2.0}
\plotone{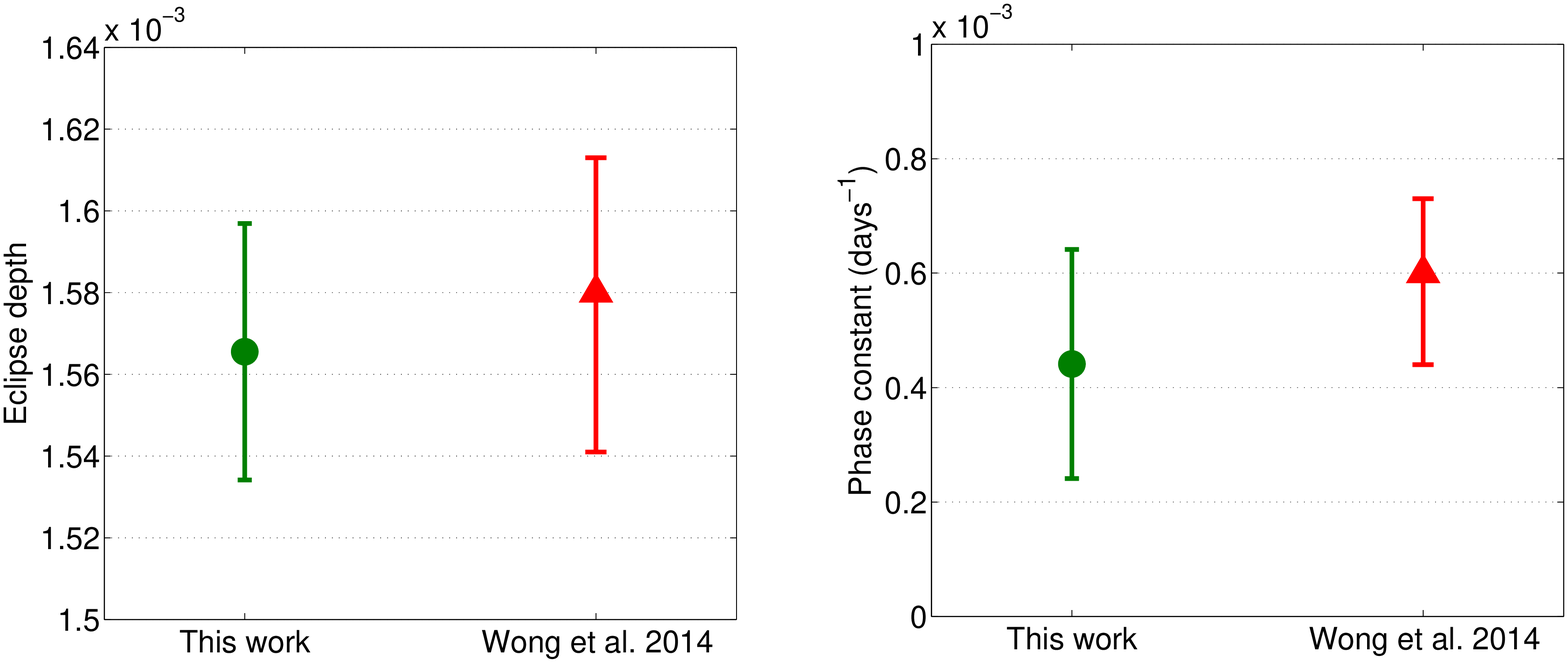}
\caption{Left panel: (green circle) best global eclipse depth estimate obtained in this work, and (red triangle) in \cite{wong14}. Right panel: the same for the global phase constant. \label{fig4}}
\end{figure*}

\section{Discussion}
\subsection{Reduced chi-squared tests}
The underlying assumption for the weighted mean to be a valid parameter estimate is that individual measurements of that parameter are normally distributed around the same mean value with variances $\sigma_i^2$, and there are no systematic errors. The reduced chi-squared can be used to test, in part, this hypothesis:
\begin{equation}
\chi_0^2 =  \frac{1}{n-k} \sum_{i=1}^{n} \frac{(x_i - \bar x)^2}{\sigma_i^2}
\end{equation}
where $x_i \pm \sigma_i$ are the individual measurements, $\bar x$ is the weighted mean value, $n=$12 is the number of measurements, and $k=$1 is the number of calculated parameters.  Ideally, if the assumption is valid, we should expect $\chi_0^2 \lesssim$1. Conventionally, the hypotesis is rejected if $\chi_0^2 > M_{n,k}$, where $M_{n,k}$ is the critical value corresponding to a probability of less than 5$\%$ for the hypothesis to be valid. We found $\chi_0^2 =$0.42 for the eclipse depth, and $\chi_0^2 =$1.0 for the phase constant, confirming the non-detection of any inter-epoch variability. $\chi_0^2 =$0.42 may suggest that the error bars for the eclipse depth are over-estimated, but this is not totally surprising, given that we actively increased them to account for potential uncorrected systematics and biases in the detrending method. Note that the reduced chi-squared tests whether the actual dispersion in the measurements is consistent within their error bars, but it is not sensitive to a uniform bias for all measurements, e.g. a constant shift. Hence, it is not sufficient alone to justify the use of the weighted mean as global estimate of a parameter. Additional tests, reported in the Appendices, show that the weighted mean result is very stable for the eclipse depth. The phase constant appears to be more dependent on certain detrending options, in particular background subtraction. In this case, the adopted weighted mean error bar of 2.0$\times$10$^{-4}$ days$^{-1}$ is a lower limit, valid under some caveats. In the worst-case scenario, the maximum error bar, calculated without scaling when combining multiple measurements, is 7$\times$10$^{-4}$ days$^{-1}$.

\subsection{Comparison with a previous analysis of the same observations}
\label{sec:wong}
Our results are consistent within 1 $\sigma$ with the ones from a previous analysis reported in \cite{wong14} (see Figures \ref{fig3} and \ref{fig4}). Our error bars are generally larger by a factor 0.8-1.5 for the eclipse depth (smaller in 1 case) and 1.0-2.0 for the phase constant compared to the ones in the literature. The factors for the weighted mean eclipse depth and phase constant are 0.9 and 1.4, respectively. Slightly larger error bars are a worthwhile trade-off for much higher objectivity, which derives from the lack of assumptions about the origin of instrument systematics and their functional forms in our detrending method. We also note that, despite the larger nominal error bars, the dispersions in our best parameter estimates are slightly smaller than the ones calculated from the results reported in \cite{wong14} (see Table \ref{tab4}).

The reduced chi-squared values inferred from their individual parameter estimates are $\chi_0^2 =$0.86 for the eclipse depth, and $\chi_0^2 =$4.3 for the phase constant. While the first $\chi_0^2$ value is consistent with the hypothesis of a constant transit depth within the quoted error bars, the second $\chi_0^2$ value indicates that the analogous hypothesis for the phase constant can be rejected (less than 0.1$\%$ probability of being true). This may suggest either that they were able to detect some astrophysical variability of the phase curve's slope, or that their individual error bars are under-estimated by a factor $\sim$2. Given that the astrophysical slope is degenerate with other instrumental trends, such as long-term position drift of the telescope and possible thermal heating, it is possible that their individual error bars do not fully accounts for these degeneracies, as the authors themselves state. If this is the case, we observe that their final error bar on the phase constant, derived by a joint fit of all eclipses, could be equally under-estimated, because it is not guaranteed that residual systematic errors cancel out over multiple observations as if they were random errors. Note that the joint fit approach is theoretically valid under the same assumptions for which the weighted mean is valid, and the two approaches are expected to lead to very similar results (we checked that this happens in this case). In conclusion, our reanalysis confirms the results reported in \cite{wong14} for the eclipse depth, and relative inter-epoch variability, but not the 4$\sigma$ detection of a non-flat phase curve's slope during the eclipse, as larger error bars are needed to account for the possible residual systematics.
\begin{table*}[!h]
\begin{center}
\caption{Weighted mean parameter results, dispersions and reduced chi squared values obtained in this paper and reported in \cite{wong14}. \label{tab4}}
\begin{tabular}{ccc}
\tableline\tableline
Eclipse depth & This work & Wong et al. 2014\\
\tableline
Best estimate & (1.57 $\pm$ 0.03)$\times$10$^{-3}$ & 1.580$_{-0.039}^{+0.033}\times$10$^{-3}$\\
Dispersion & 7.2$\times$10$^{-5}$ & 8.4$\times$10$^{-5}$\\
$\chi_0^2$ & 0.42 & 0.86\\
\tableline\tableline
Phase constant (days$^{-1}$) & This work & Wong et al. 2014\\
\tableline
Best estimate & (4.4 $\pm$ 2.0)$\times$10$^{-4}$ & 6.0$_{-1.6}^{+1.3}\times$10$^{-4}$\\
Dispersion & 7.0$\times$10$^{-4}$ & 11.3$\times$10$^{-4}$\\
$\chi_0^2$ & 1.0 & 4.3\\
\tableline
\end{tabular}
\end{center}
\end{table*}

\section{Conclusions}
We have applied a blind signal-source separation method to analyze twelve photometric observations of the eclipse of the exoplanet XO3b obtained with Warm \textit{Spitzer}/IRAC at 4.5 $\mu$m. The method is an evolution of the pixel-ICA method proposed and successfully used by our team to analyze real and synthetic transit observations. By adding a wavelet transform of the time series, we extend the applicability of pixel-ICA to detrend low-S/N observations with instrumental systematics stronger than the astrophysical signal.
Wavelet pixel-ICA results are consistent within 1 $\sigma$ with results reported in the literature. They also have smaller dispersions in the eclipse parameters measurements, even including the most recent results that appeared on the arXiv while this paper was under review. While the individual error bars are usually more conservative, as they fully accounts for the possible uncertainties, the final error bar on the eclipse depth is equal or smaller than the ones obtained with other methods discussed in the literature.

No significant inter-epoch variations are detected over twelve repeated observations in 3 years interval. This is convincing evidence that, with appropriate data detrending methods, transit and eclipse measurements based on \textit{Spitzer}/IRAC observations can achieve this level of precision and reproducibility, and therefore are useful to characterize the atmospheres of exoplanets. Also, the lack of any detectable astrophysical variability, for the XO3b system, allows to combine multiple observations to increase the accuracy in stellar and planetary parameters.

\acknowledgments

This work is based on observations made with the Spitzer Space Telescope, which is operated by the Jet Propulsion Laboratory, California Institute of Technology under a contract with NASA.
This research has made use of the NASA/ IPAC Infrared Science Archive, which is operated by the Jet Propulsion Laboratory, California Institute of Technology, under contract with the National Aeronautics and Space Administration.
G. Morello acknowledges UCL Perren/Impact scholarship (CJ4M/CJ0T) and the Royal Astronomical Society. I. P. Waldmann is funded by the European Research Council Grant ``Exolights'', STFC and RCUK. G. Tinetti is funded by the Royal Society. 

\appendix

\section{Testing the robustness of results}
\subsection{The effect of binning}
\label{app_bin}
Given the large number of frames (14,912) for each observation, binning data is very useful to decrease the computational time needed to run the MCMCs for the eclipse parameters and scaling coefficients for the independent components (see Section \ref{MCMC}). Some authors suggest that an optimal choice of the binning size can be useful to reduce the noise on the timescale of interest \citep{dem15, kam15}, provided that the theoretical curve is similarly binned as necessary to eliminate bias, and the bin size is not too long to cause significant loss of astrophysical information \citep{kip10}.

An additional choice for the pixel-ICA algorithm is whether to bin the pixel time series prior the ICA separation, or to bin the independent components extracted from unbinned pixel time series. We found that the two options are almost equivalent, as the eclipse signals obtained after removing the systematic components from the raw light-curve are identical (discrepancies 1-2  order of magnitudes smaller than the fitting residuals), except in cases for which the unbinned ICA separation fails to retrieve an eclipse component. It is worth to note that for the unbinned case, the amplitude of the total noise plus systematics is higher than the the eclipse depth. Thus we decided to bin the individual pixel-light-curves prior ICA retrieval.

\begin{figure}[!h]
\epsscale{1.0}
\plotone{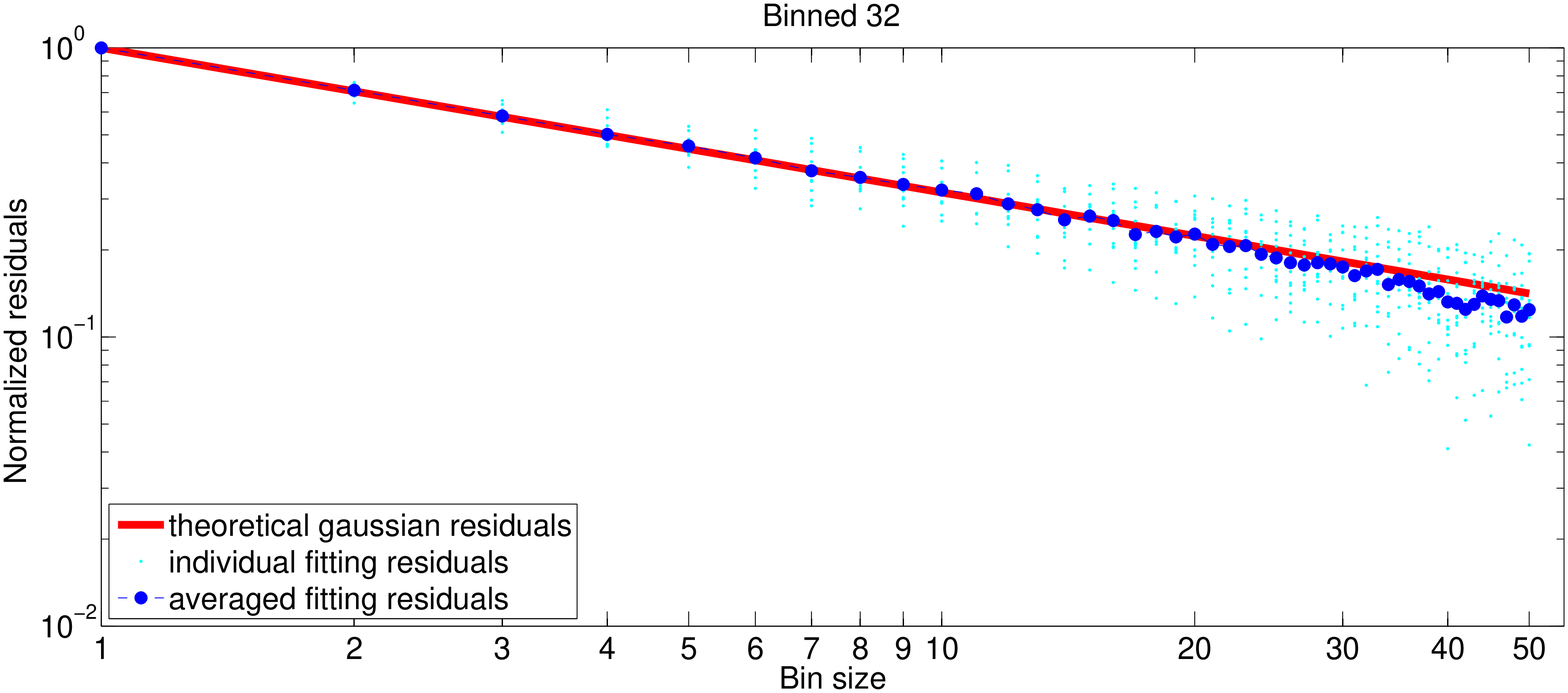}
\plotone{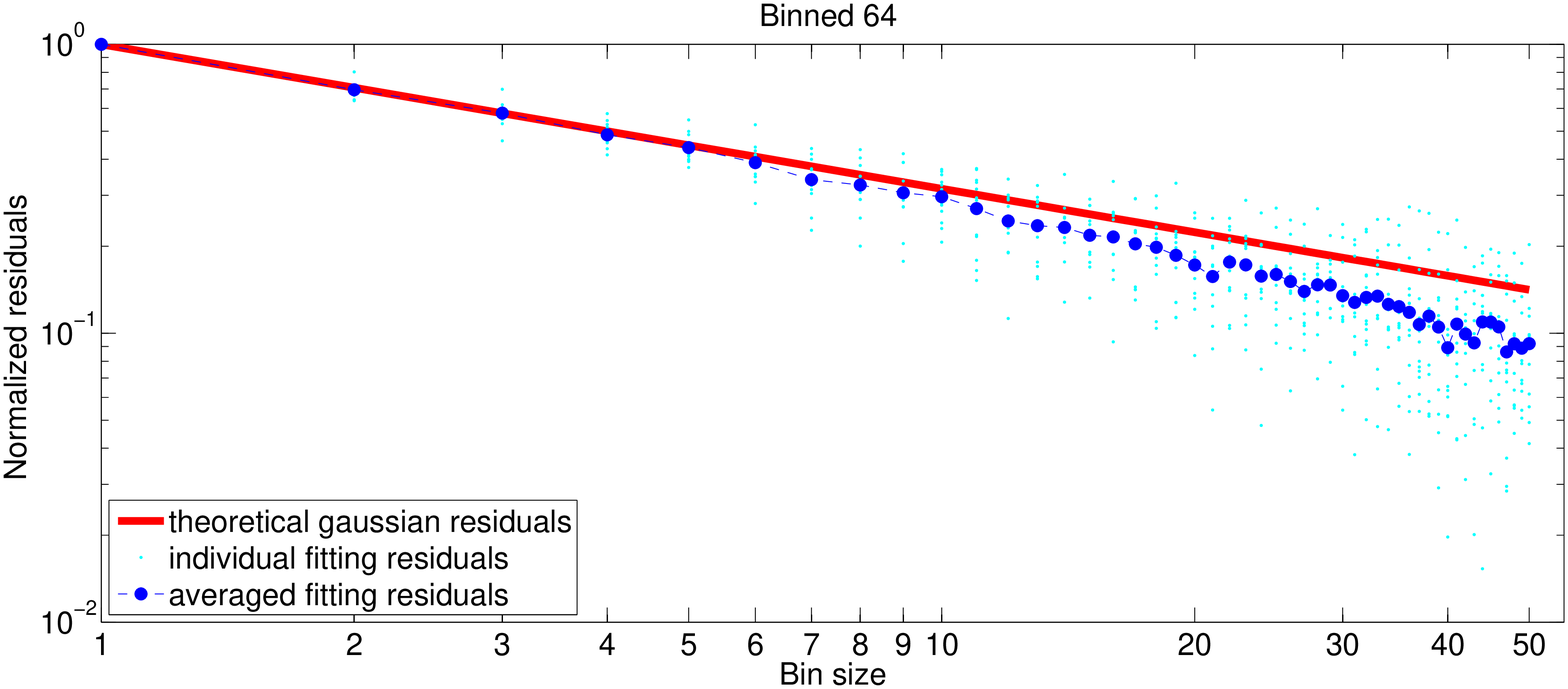}
\caption{Top panel: Root mean square of residuals as a function of bin size for (red line) gaussian white noise, (cyan dots) individual observations analyzed with bin size of 32 frames, and (blue circles) values averaged over the twelve observations. Bottom panel: the same for individual observations analyzed with bin size of 64 frames. \label{fig5}}
\end{figure}

\begin{figure}[!h]
\epsscale{1.0}
\plotone{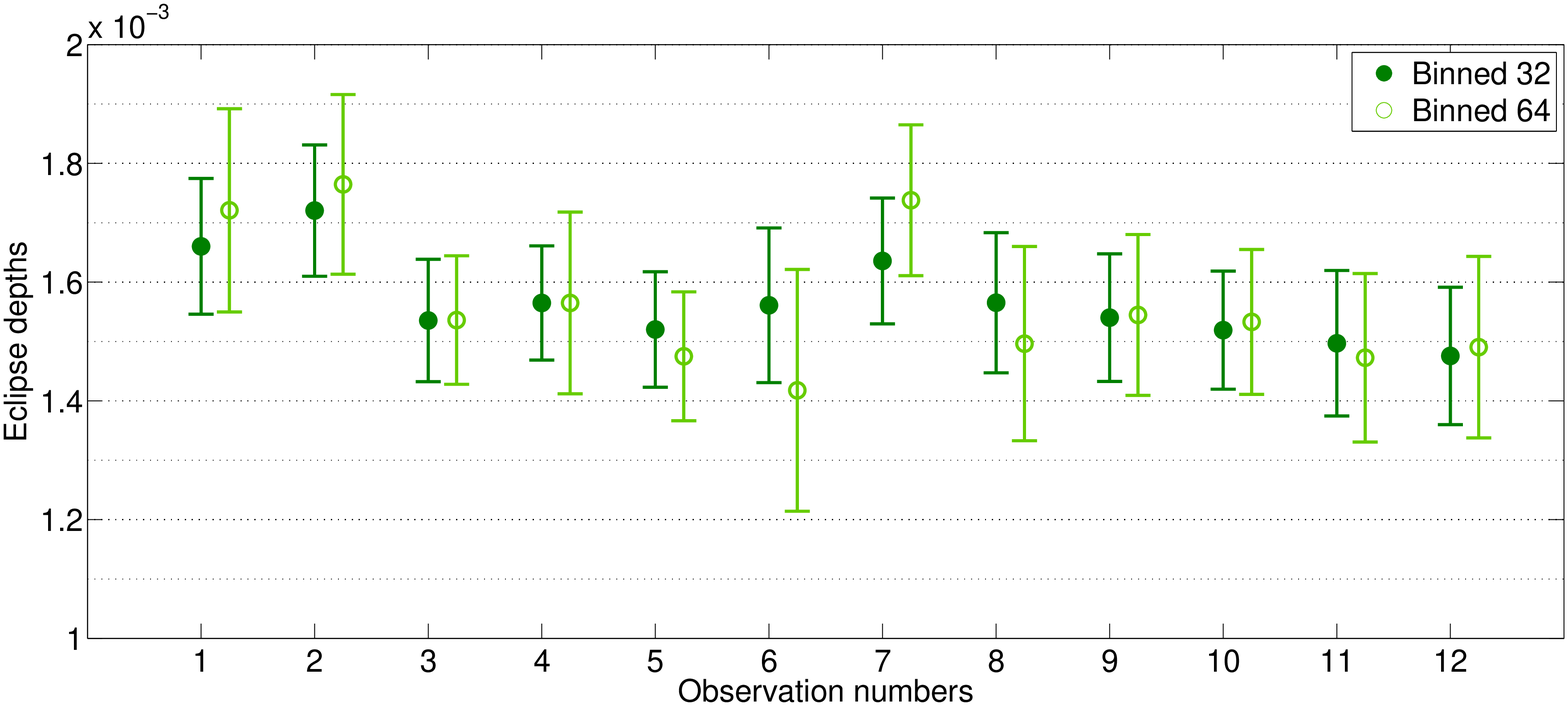}
\plotone{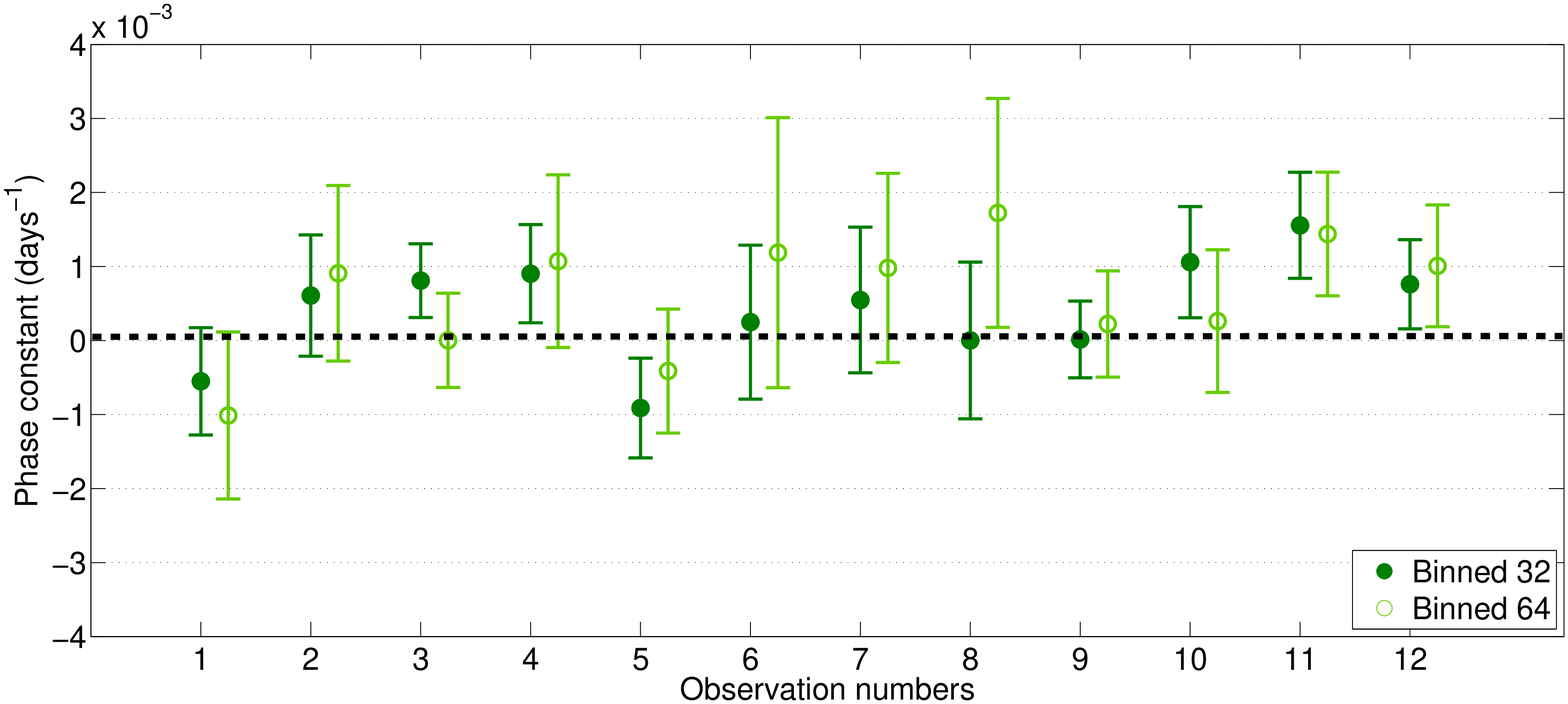}
\caption{Top panel: individual eclipse depth measurements obtained with 5$\times$5 array, background subtraction, and (dark green, full circles) binning over 32 frames, and (light green, empty circles) binning over 64 frames. Bottom panel: the same for individual measurements of the phase constant.\label{fig6}}
\end{figure}

We compare the results obtained for all the observations with bin sizes of 32 and 64 frames, i.e. 64 and 128 s, respectively. First, we test the gaussianity of fitting residuals by calculating their root mean square (rms) as a function of the bin size, $b$. If fitting residuals are white noise, the rms would scale as $1/\sqrt{b}$. Figure \ref{fig5} shows that, in both cases, the rms of fitting residuals slightly deviates from the expected behaviour of white noise. Those deviations are smaller for the analysis with bin size of 32 frames.
The parameter results obtained with the two binning choices are all consistent within 0.5 $\sigma$, and in average within 0.16 $\sigma$ (see also Figures \ref{fig6} and \ref{fig12}). Both the error bars and the overall dispersions are smaller for the cases with bin size of 32 frames by factors of $\sim$1.4. We consider the results obtained with bin size of 32 frames as our best results.

\clearpage

\subsection{Specific wavelet ICA options}
\label{app_dwt}
The DWT of a time series is specified by the choice of a mother wavelet and the number of levels (see Section \ref{DWT}). We adopted a Daubechies-4 mother wavelet, and one-level decomposition. For a few observations we tested multiple choices of the mother wavelet, among the Daubechies, Biorthogonal, Symlets and Coiflets families, and number of decomposition levels. We refer to \cite{dau92, pw00} for details about the wavelet properties. We found that different choices of the mother wavelet are not significant, e.g. discrepancies in the eclipse signals are 1-2 orders of magnitudes smaller than the fitting residuals, while level decompositions higher than 1 usually appear to make impossible for ICA to retrieve the eclipse. The difficulties with higher-level DWTs may arise from sub-sampling the average coefficients, and the fact that some of the low-frequency non-gaussian components may be smeared over higher levels of detail coefficients.

\subsection{About background subtraction}
\label{app_back}
Uncorrected background may bias the normalized amplitude of the eclipse depth, as well as the phase curve's slope, if background is not constant over time. The typical morphology of a background time series is either a constant function or a slow monotonic drift. The lack of a distinct temporal structure and non-gaussianity makes it difficult to disentangle with ICA, as well as other statistical methods. For this reason, we performed ad hoc background subtractions before ICA detrending (see Section \ref{back}). In this Section, we discuss the impact of this step in the analyses, by comparing results obtained with and without background subtraction.  These are also used to infer the maximum parameter errors that can be caused by an inappropriate background correction.

\begin{figure}[!h]
\epsscale{1.0}
\plotone{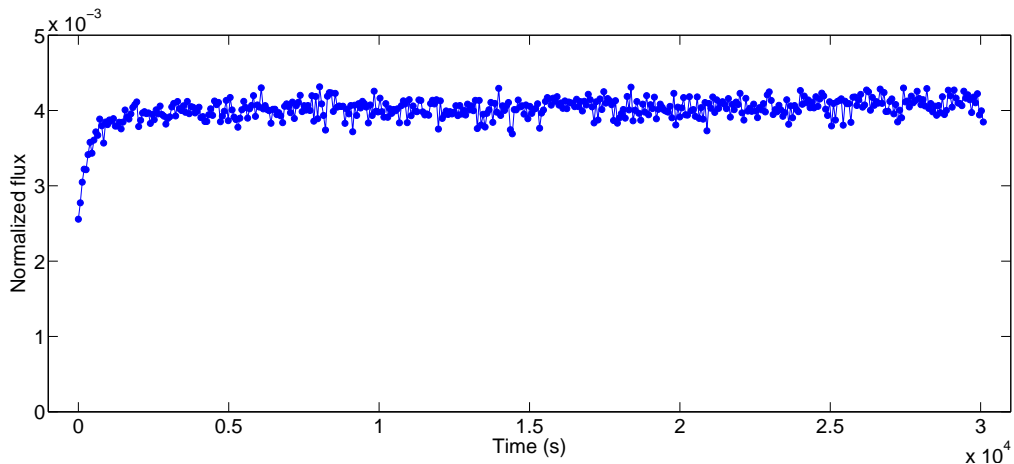}
\caption{Example of background time series binned over 32 frames.\label{fig7}}
\end{figure}

Figure \ref{fig7} shows an example of background time series estimated for one observation. The measured mean background level slightly varies from one observation to the other, but it is always less than 0.6$\%$ of the total flux from the system, hence potentially affecting the eclipse depth by less than 10$^{-5}$, well below the error bars. Background is also not constant during one observation: it has a small ramp for the first $\sim$20 minutes (except for the eclipses extracted from longer observations), then continues to slowly increase.

\begin{figure}[!h]
\epsscale{1.0}
\plotone{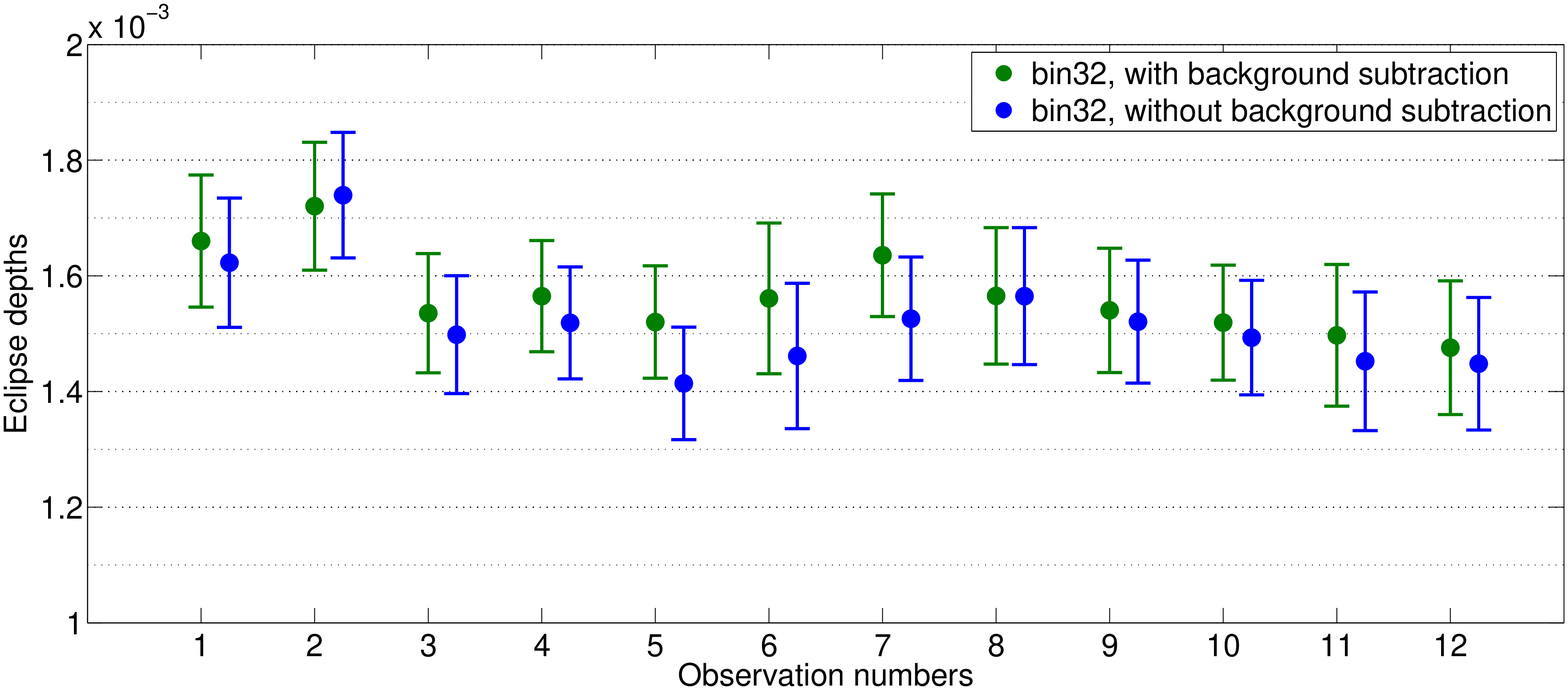}
\plotone{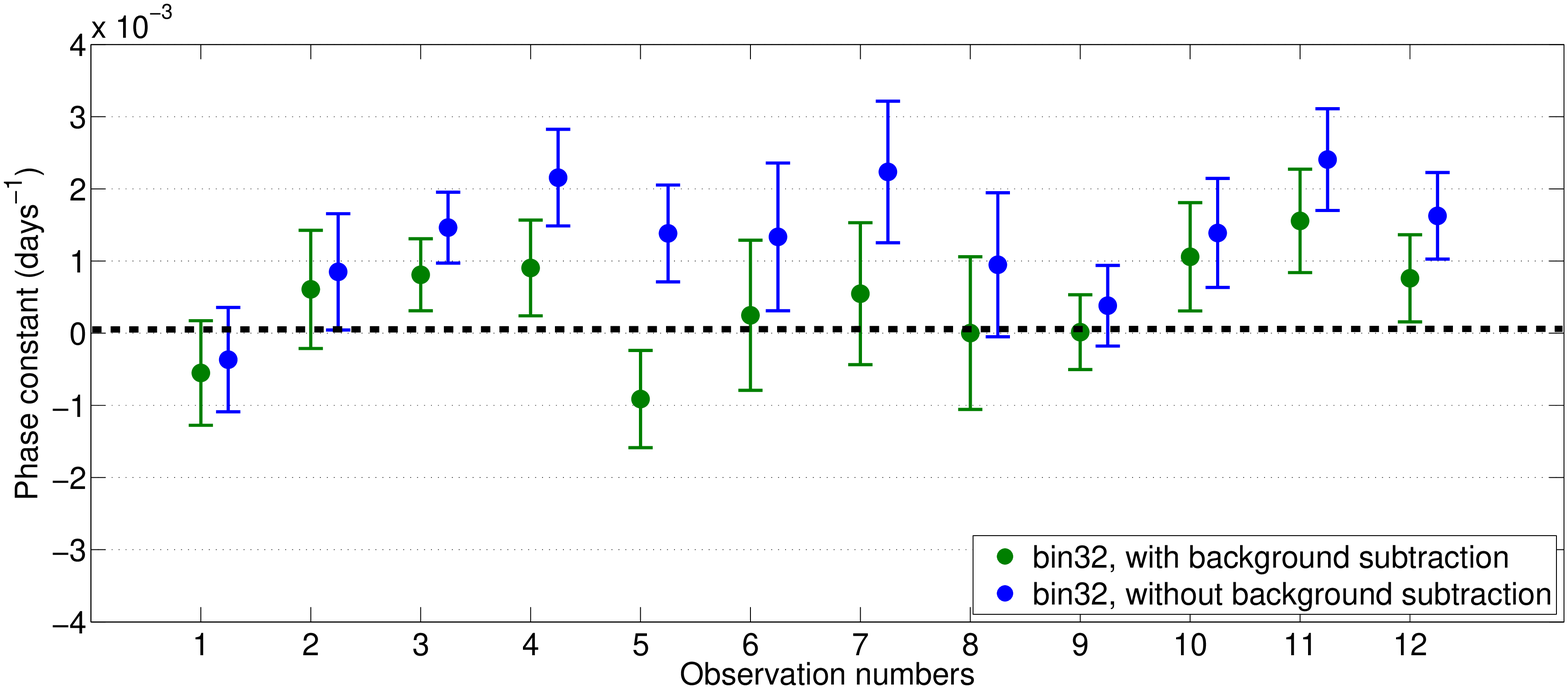}
\caption{Top panel: individual eclipse depth measurements obtained with 5$\times$5 array, time series binned over 32 frames (green circles) with background subtraction, and (blue circles) without background subtraction. Bottom panel: the same for individual measurements of the phase constant.\label{fig8}}
\end{figure}

Figure \ref{fig8} shows the parameter results obtained with and without background subtraction. The greatest discrepancies are observed for the phase constant, which is systematically higher by 2-20$\times$10$^{-4}$ days$^{-1}$ for the cases without background subtraction, suggesting the presence of uncorrected systematics. It is quite remarkable that our individual error bars automatically account for those systematics, but, given their non-random nature, the weighted mean error bars for the phase constant cannot be compared. It is difficult to prove the superiority of the results obtained with background subtraction over the others, as the residuals between the full models and the relevant raw light-curves (see Section \ref{MCMC}) are very similar for the two cases, e.g. similar amplitudes and time correlations, leading to similar error bars. Slightly smaller parameter dispersions are obtained with background subtraction rather than without, in particular 7$\times$10$^{-5}$ vs 9$\times$10$^{-5}$ for the eclipse depth, and 7$\times$10$^{-4}$ vs 8$\times$10$^{-4}$ days$^{-1}$ for the phase constant. The higher mean value of the phase constant obtained without background subtraction, i.e. $\sim$13$\times$10$^{-4}$ days$^{-1}$, appears to be less likely, as it would require a higher than expected increase in the atmospheric temperature due to stellar irradiation during the eclipse, and/or strong horizontal disomogeneities either in temperature, chemical composition and/or clouds \citep{ca11, kat13, agu14}. If we assume that systematics are removed with background subtraction, we can take the weigthed mean as best estimate for the phase constant.

The discrepancies between eclipse depth measurements with and without background subtraction are in the range 10$^{-5}$-10$^{-4}$, and, in average, the eclipse depth is smaller by $\sim$4$\times$10$^{-5}$ for the case without background subtraction. Although this is more than the 10$^{-5}$ difference expected from the mean background level relative to the mean stellar flux (see discussion in the previous paragraph), in this case, the two weighted means are consistent within 0.5 $\sigma$. We found that the main effect of background on the eclipse depth is due to correlations between the measured eclipse depth at the eclipse center and the phase constant, in terms of Pearson correlation coefficients between the relevant MCMCs. The details of this study are beyond the scope of this paper.

\clearpage

\subsection{Using different arrays of pixels}
In previous studies, we found that, for high-S/N transit observations, pixel-ICA performances are only slightly dependent on the choice of the array \citep{mor14,mor15}. The 3$\times$3 array is too narrow compared to the Spitzer/IRAC Point Spread Function, and for this reason, it is not photometrically stable (although it usually leads to consistent results). Larger arrays are more photometrically stable, but they include more noise.  Also, noisier pixel-light-curves, appear to increase the uncertainties in the ICA decomposition ($\sigma_{ICA}$, see Section \ref{error_bars}), so that the smallest error bars were usually obtained using the 5$\times$5 array.

\begin{figure}[!h]
\epsscale{1.0}
\plotone{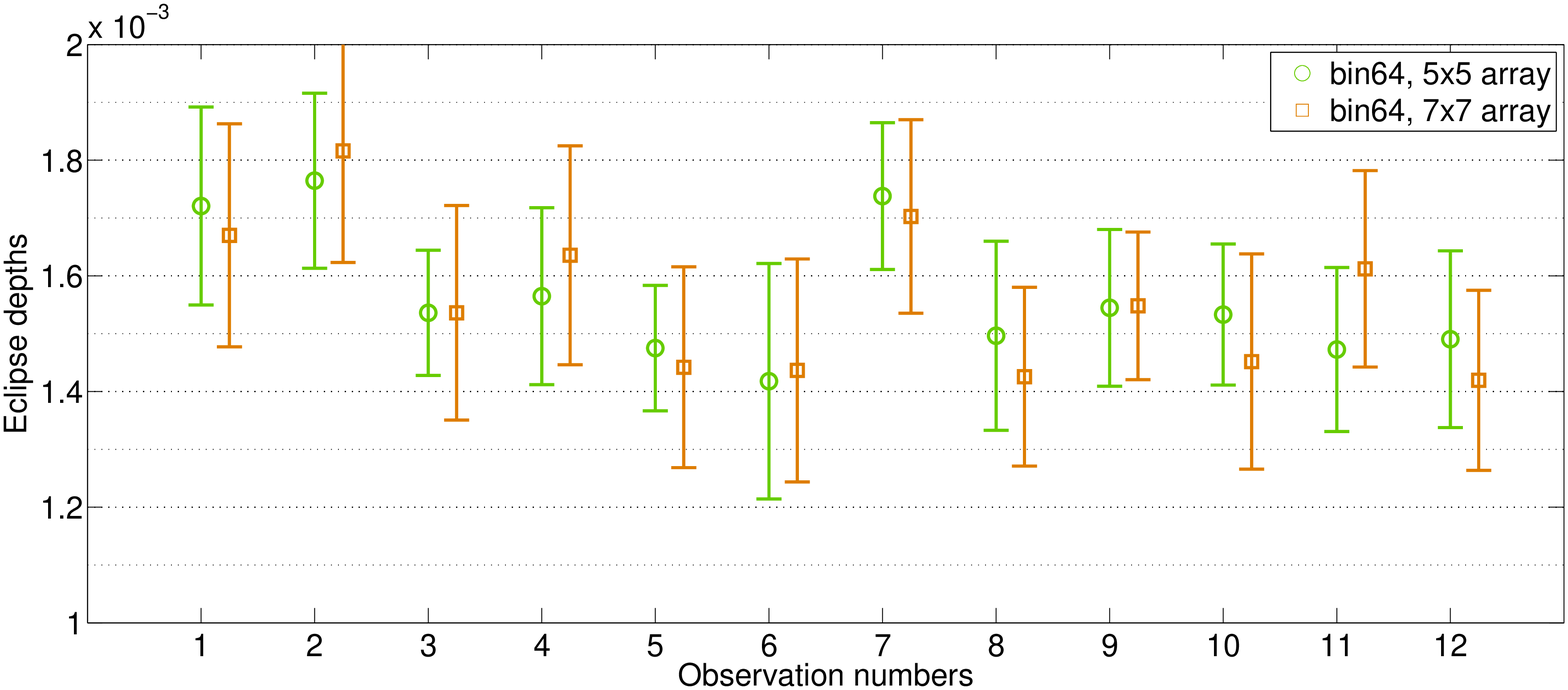}
\plotone{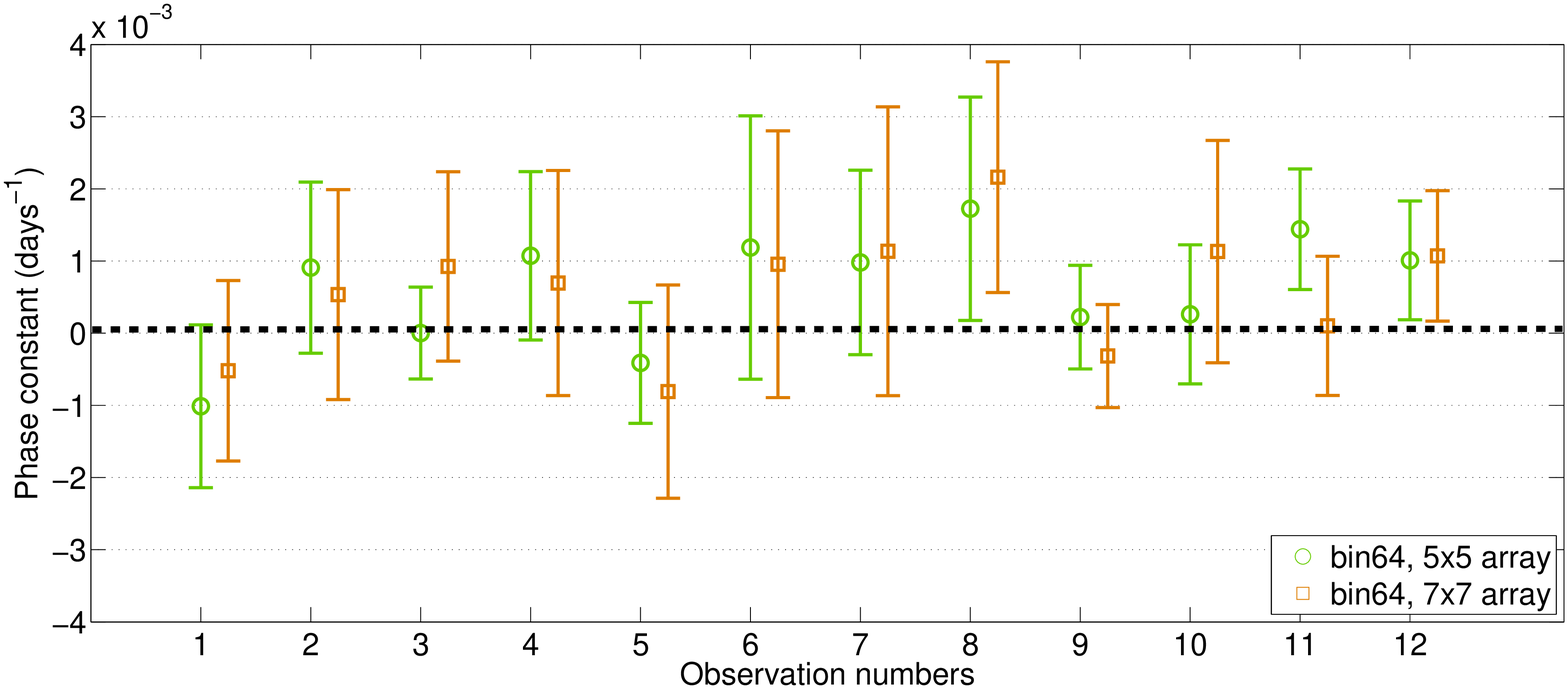}
\caption{Top panel: individual eclipse depth measurements obtained with (green circles) 5$\times$5 array, and (brown squares) 7$\times$7 array;  both are obtained with time series binned over 64 frames, and background subtraction. Bottom panel: the same for individual measurements of the phase constant.\label{fig9}}
\end{figure}

In this work, we analyzed all datasets with two different choices of the pixel-array, i.e. 5$\times$5 and 7$\times$7. For the analyses with the 7$\times$7 array, we only adopted the 64 frames bin size, for which the MCMC fitting is faster. Figure \ref{fig9} compares the results obtained with the two different arrays and the same bin size. The two sets of results are consistent well within 1 $\sigma$, but error bars obtained with the 7$\times$7 array are 1-1.5 times larger than the ones obtained with the 5$\times$5 array, despite the residuals in the fits are similar and often smaller for the 7$\times$7 cases. This confirms the conclusions obtained from previous analyses, in particular:
\begin{enumerate}
\item the 5$\times$5 array leads to smaller error bars than other squared arrays of pixels;
\item parameter results from different arrays are consistent well within 1 $\sigma$.
\end{enumerate}
In this case, the choice of a less optimal array increases the error bars more significantly than in our previous analyses, most likely because of the lower S/N of the observations analyzed here.

\begin{figure}[!h]
\epsscale{1.0}
\plotone{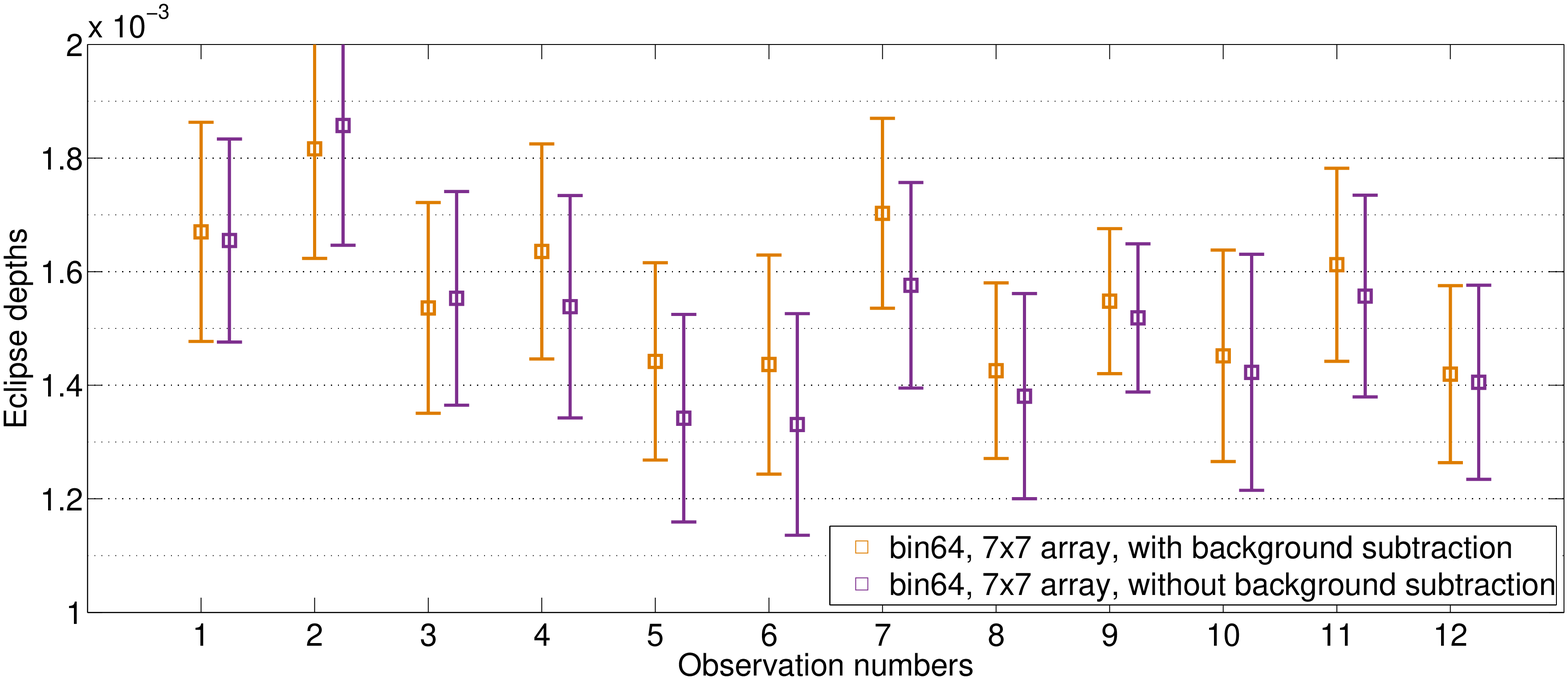}
\plotone{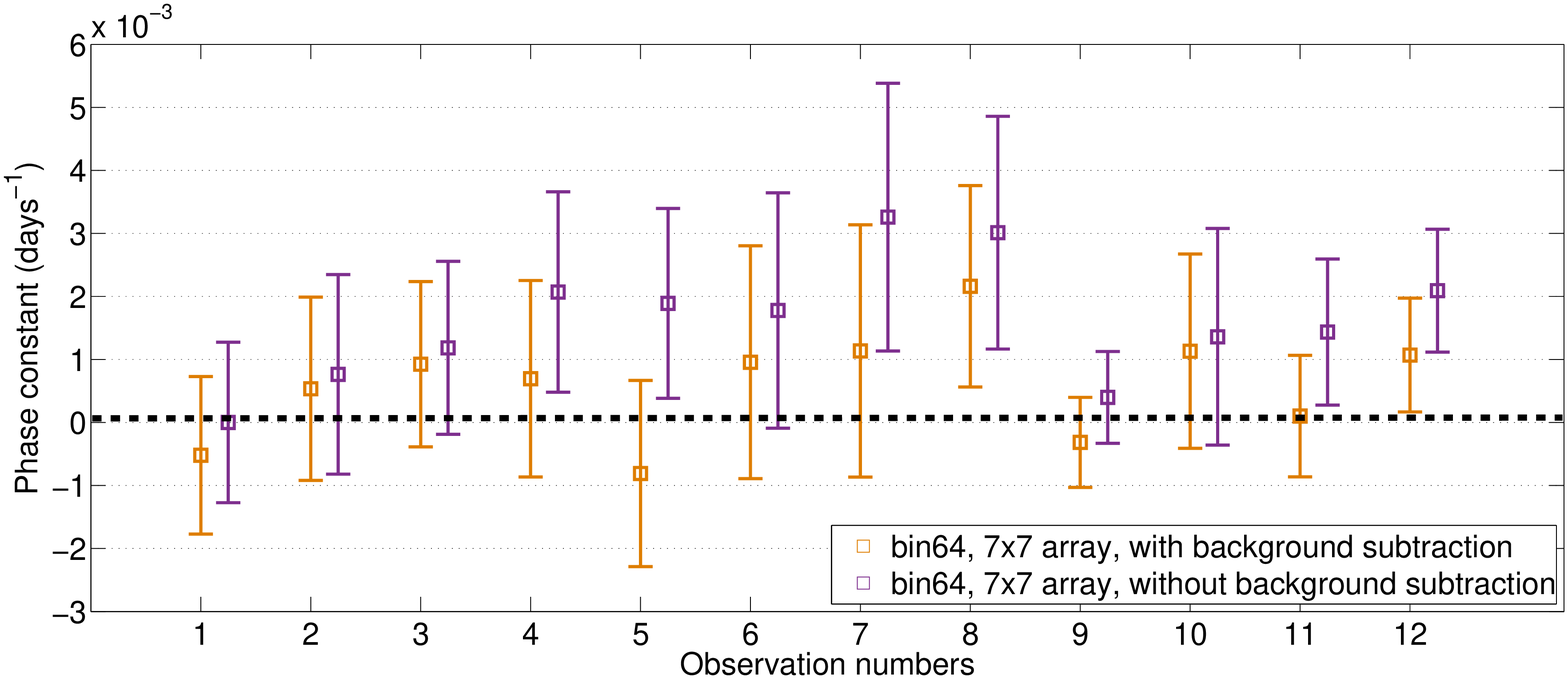}
\caption{Top panel: individual eclipse depth measurements obtained with 7$\times$7 array, time series binned over 64 frames (brown squares) with background subtraction, and (purple squares) without background subtraction. Bottom panel: the same for individual measurements of the phase constant.\label{fig10}}
\end{figure}

Figure \ref{fig10} compares the results obtained with and without background subtraction for the case of 7$\times$7 array. The same considerations discussed in Appendix \ref{app_back} for the 5$\times$5 array are valid for the 7$\times$7 array.

\clearpage

\subsection{Breaking degeneracy with the phase constant}
\label{app_zeroslope}
At the end of Appendix \ref{app_back} we revealed the existence of a correlation between the eclipse depth and phase constant parameter, which is stronger for data with a higher slope, such as the cases without background subtraction.
This may affect the eclipse depth estimate in case of residual systematics with a slope, e.g. uncorrected background. In our analyses, the maximum bias on the eclipse depth due to correlations with this kind of systematics is $\sim$4$\times$10$^{-5}$, provided the systematic slope is not larger than our individual error bars.

\begin{figure}[!h]
\epsscale{1.0}
\plotone{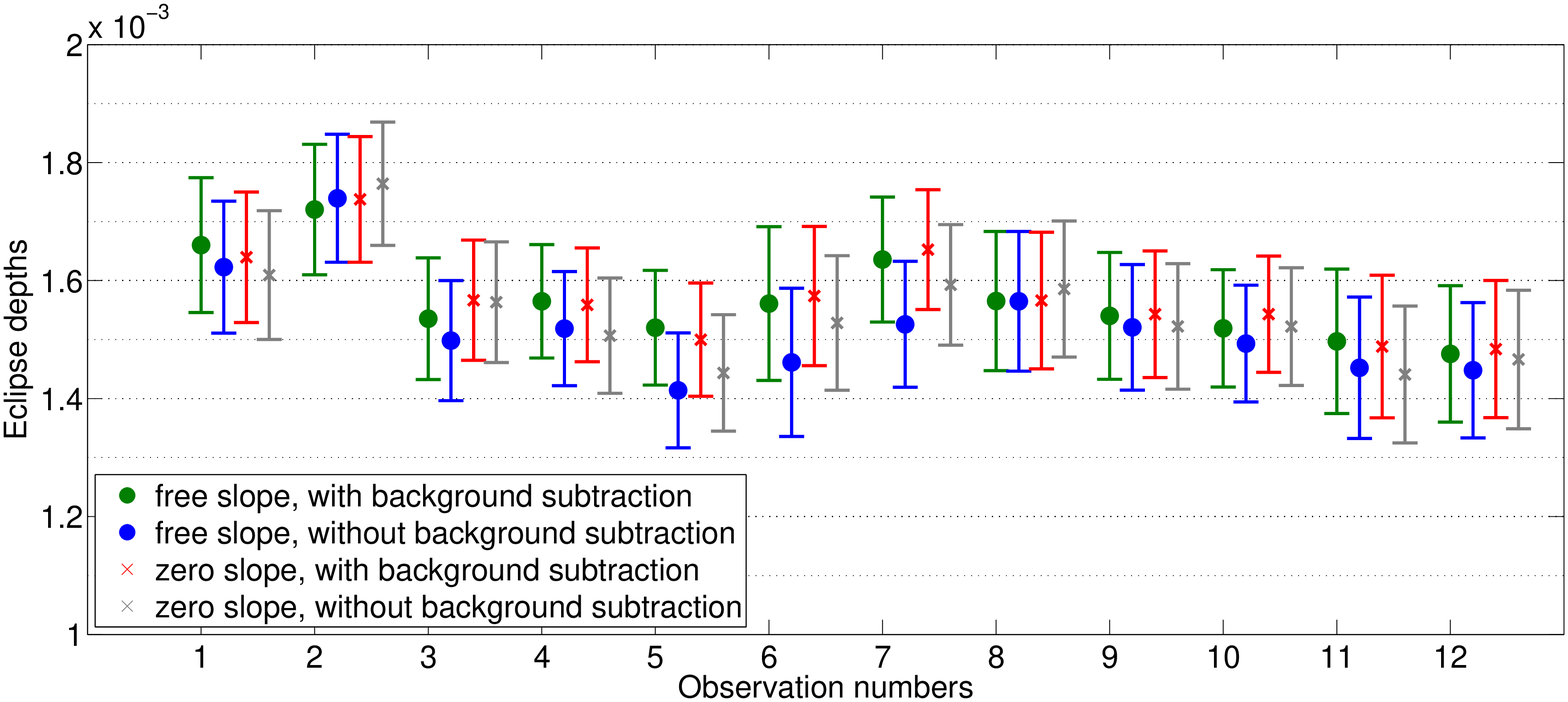}
\caption{Individual eclipse depth measurements obtained with 5$\times$5 array, time series binned over 32 frames (green circles) with fitted phase constant and background subtraction, (blue circles) with fitted phase constant and without background subtraction, (red `x') with zero phase constant and background subtraction, and (grey `x') with zero phase constant and without background subtraction.\label{fig11}}
\end{figure}

Here we test the consequences of adopting zero phase constant while fitting for the eclipse depth. This is equivalent to the assumption that the phase curve is flat before and after the eclipse, and the slope is entirely due to instrumental effects. Compared to the cases with the phase constant as free parameter, fitting residuals with zero phase constant are not significantly larger. Figure \ref{fig11} compares results for the eclipse depth with free and zero phase constant, with and without background subtraction. The weighted mean results are reported in Figure \ref{fig12}.
We note that differences in eclipse depth measurements with and without background subtraction are smaller for the zero phase constant models, suggesting that eclipse depth measured with zero phase constant models are less affected by residual systematics with a slope. Free phase constant models are valid in a more general context, as, differently from the zero phase constant models, they approximate planet's flux variability during the observation. The consistency between the results obtained with the two classes of models indicates that, in this case, planet's flux variability in the proximity of the eclipse is smaller than the error bars.

\subsection{Robustness of the eclipse depth measurement}
Figure \ref{fig12} reports the weighted mean eclipse depth estimated for all the tests discussed in the Appendices. Note that they are mutually consistent at the 0.5 $\sigma$ level, and the range is 6$\times$10$^{-5}$.
\begin{figure}[!h]
\epsscale{1.0}
\plotone{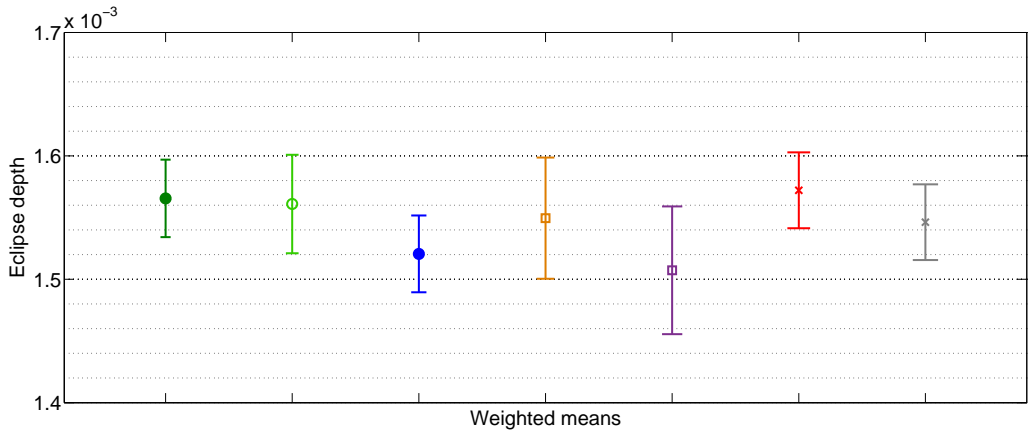}
\caption{Weighted mean eclipse depth values obtained with different tests: 
(dark green, full circle) 5$\times$5 array, with background subtraction, bin over 32 frames;
(light green, empty circle) 5$\times$5 array, with background subtraction, bin over 64 frames;
(blue, full circle) 5$\times$5 array, without background subtraction, bin over 32 frames;
(brown, empty square) 7$\times$7 array, with background subtraction, bin over 64 frames;
(purple, empty square) 7$\times$7 array, without background subtraction, bin over 64 frames;
(red `x') 5$\times$5 array, with background subtraction, bin over 32 frames, and zero phase constant;
(grey `x') 5$\times$5 array, without background subtraction, bin over 32 frames, and zero phase constant.
\label{fig12}}
\end{figure}


\clearpage

\end{document}